\documentclass[12pt,preprint]{aastex}

\newcommand{\teff}{T$_{\rm eff}$} 

\begin{document}

\title{Magnesium isotopic abundance ratios in cool stars}

\author{David Yong and David L. Lambert}
\affil{Department of Astronomy, University of Texas, Austin, TX 78712}
\email{tofu,dll@astro.as.utexas.edu}

\and

\author{Inese I.\ Ivans\altaffilmark{1}}

\affil{Astronomy Department, California Institute of Technology, Pasadena, CA 91125}
\email{iii@astro.caltech.edu}

\altaffiltext{1}{Hubble Fellow}

\begin{abstract}

From high resolution spectra of 61 cool dwarfs and giants, Mg isotopic abundance
ratios $^{24}$Mg:$^{25}$Mg:$^{26}$Mg are derived from spectral synthesis of the MgH A-X lines
near 5140\AA.  Our sample spans the range $-2.5 \le$ [Fe/H] $\le 0.1$,
including the first measurements of Mg isotope ratios in stars with metallicities
below [Fe/H]=$-$2.0.  We confirm the decrease in 
$^{25,26}$Mg/$^{24}$Mg with decreasing [Fe/H] as predicted by
recent models of Galactic chemical evolution where the Mg isotopes 
are produced in massive stars.  A subset of kinematically identified thin disk stars
have Mg isotope ratios in excellent agreement with the predictions.  Within the
measurement uncertainties, these thin disk stars show no scatter about the predictions.
Several of our stars are likely members
of the thick disk and their high Mg isotopic ratios may 
reflect the nucleosynthetic history of the thick disk
which is distinct from the predictions for, and observations of, the thin disk.
For thick disk and halo stars we find a scatter in 
$^{25,26}$Mg/$^{24}$Mg exceeding our measurement
uncertainties and increasing with increasing metallicity.
Our data suggest that 
an additional source of $^{25}$Mg and $^{26}$Mg is required.  Intermediate 
mass asymptotic giant branch stars are likely candidates.  

\end{abstract}

\keywords{stars: abundances --  subdwarfs -- Galaxy: abundances -- Galaxy: evolution}

\section{Introduction}
\label{sec:intro}

Studies of the chemical evolution of the Galaxy seek a full
understanding of the chemical composition of objects -- past and present --
throughout the Galaxy.  Much of the observational data have been provided by stars
now in the solar neighborhood whose ages and places of birth may differ
considerably from our Sun's present location and age.  
Elemental abundances observed in large samples of disk stars (e.g., \citealt{bdp93,bdp03}),
extremely metal-poor stars (e.g., \citealt{mcwilliam95,ryan96}), and the general
halo population (e.g., \citealt{fulbright00})
have provided insights
into the cycle of star formation, evolution, and death that has controlled
the composition of gas in the Galaxy.
Theoretical efforts (including \citealt{timmes95}, \citealt{goswami00}, and \citealt{alc01})
present predictions of the evolution of elements from carbon
to zinc where the basic ingredients in these models 
of Galactic chemical evolution include stellar
yields, the initial mass function, and the star formation rate.  The success
of a model can be gauged by how accurately the predictions match the observational 
data.  In turn, the observations can be used to constrain the models.
The magnesium isotopes present a rare opportunity where the
evolution of isotopic abundances can be measured to test directly models
of Galactic chemical evolution.

The stable isotopes of Mg consist of the dominant $^{24}$Mg and the 
neutron-rich minor isotopes $^{25}$Mg and $^{26}$Mg.  
Massive stars
produce $^{24}$Mg in their carbon and neon burning shells before
their deaths as Type II supernovae \citep{arnett85,thielemann85}.
Helium burning is responsible for the synthesis of the less abundant isotopes via 
$^{22}$Ne($\alpha,n)^{25}$Mg($n,\gamma)^{26}$Mg.  The production of 
the heavier Mg isotopes relies upon the abundance of $^{22}$Ne
which is set primarily by the initial abundances of C, N, and O.
A consequence of the H-burning CNO cycle is that 
in equilibrium, $^{14}$N is the dominant nuclei
whose abundance is essentially equal to the initial abundances of
C+N+O.  Successive alpha captures on $^{14}$N can then produce $^{22}$Ne, the
seed from which $^{25}$Mg and $^{26}$Mg are synthesized.  Therefore, 
the yields of $^{25}$Mg and $^{26}$Mg are predicted to increase
with increasing metallicity.  Yields calculated from massive stars \citep{woosley95}
indeed verify the increase of the neutron-rich Mg isotopes with increasing
metallicity.

Beginning with \citet{boesgaard68}, the relative abundances of the
Mg isotopes have been measured in stars from analyses
of MgH lines.  
\citet{tl80} provided the first evidence that metal-poor
stars possess low ratios of $^{25}$Mg/$^{24}$Mg and $^{26}$Mg/$^{24}$Mg 
from an analysis of the subdwarf Gmb 1830.  Later studies of the stellar
Mg isotopes include those conducted by 
\citet{barbuy85}, \citet{lm86}, \citet{bpp87}, and \citet{ml88}.  
Recently, \citet{gl2000} measured 
the ratio $^{24}$Mg:$^{25}$Mg:$^{26}$Mg to [Fe/H] as low as 
$\simeq-1.5$.  Gay \& Lambert found
reasonable agreement between the measured isotopic ratios and predictions
from the \citet{timmes95} model of the solar neighborhood 
in which the Mg isotopes were produced by
massive stars.  Gay \& Lambert also demonstrated that some stars
show convincing and unusual excesses of the 
heavier Mg isotopes relative to other stars of the same [Fe/H].  In 
several of these unusual cases, a $s$-process enrichment was
also evident.  This suggests that these peculiar compositions were a direct
consequence of contamination
of the star (or the star's natal cloud) by ejecta from intermediate mass 
asymptotic giant branch (AGB) stars.  Sufficiently massive AGB
stars experience thermal pulses (He shell flashes) leading to the reaction
$^{22}$Ne($\alpha,n)^{25}$Mg (e.g., \citealt{iben75} and 
\citealt{forestini97}).  The neutrons released from this reaction
can then enrich the envelope in $s$-process elements along with 
$^{25}$Mg and $^{26}$Mg.  

\citet{gl2000} measured Mg isotopic ratios in 20 stars.  Prior to Gay \& Lambert,
Mg isotopic ratios had been measured in about 20 stars.
In this study, we report Mg isotopic abundance ratios for 61 stars.  Our
measurements extend in metallicity down to [Fe/H]=$-$2.5 in order to 
investigate the evolution of the Mg isotopic ratios at low metallicities.  

\section{Observations and data reduction}
\label{sec:data}

A dedicated search for cool subdwarfs was conducted to provide
suitable targets to extend the Mg isotopic measurements 
below [Fe/H]$\simeq -1.5$; the stars (dwarfs and giants) known to have
[Fe/H]$<-1.5$ are all too warm to provide MgH lines of adequate
strength to measure the isotopic ratios.  For the selection criteria,
analysis, and further details regarding the search for cool
subdwarfs see \citet{search} (hereafter Paper I).  For [Fe/H] $<-1.5$, the neutron-rich
isotopes are only expected to provide a small contribution to the total Mg abundance,
$^{25,26}$Mg/$^{24}$Mg $< 0.05$.  The target stars should therefore have strong MgH lines
in order to measure the neutron-rich minor isotopes.
A useful feature of the MgH molecule
is that the strength of the MgH lines does
not strongly depend on metallicity.  \citet{cottrell78} showed that for sufficiently
cool stars, a decrease in the metal abundance will weaken the atomic
lines whilst the MgH lines remain strong.  

The stars listed in Table \ref{tab:param} were observed at McDonald Observatory 
on the 2.7m Harlan J. Smith telescope
between November 1999 and April 2002.  The data were obtained
using the cross-dispersed echelle spectrometer \citep{tull95}
at the coud\'{e}~f/32.5 focus with a resolving power of 
$R \equiv \lambda/\Delta\lambda = 60,000$ (4 stars were observed with a 
resolving power of about $R = 35,000$).  The detector was a 
Tektronix CCD with 24 $\mu{\rm m}^2$ pixels
in a $2048 \times 2048$ format.  This setting provided
spectral coverage from 3800\AA~to 
8900\AA~with gaps between the orders beyond 5800\AA.  When
necessary, multiple 20-30 minute exposures were co-added to increase
the signal-to-noise ratio (S/N).
Although varying from star to star, the typical S/N
of the extracted one-dimensional spectra was 90 per pixel at 5140\AA.
One dimensional wavelength calibrated normalized spectra
were extracted in the standard way using the
IRAF\footnote{IRAF is distributed by the National Optical Astronomy Observatories,
which are operated by the Association of Universities for Research
in Astronomy, Inc., under cooperative agreement with the National
Science Foundation.} package of programs.  

The stellar parameters and uncertainties were derived in Paper I and the 
procedure will be repeated briefly here.  The equivalent widths
of Fe\,{\sc i} and Fe\,{\sc ii} lines were measured using routines in
IRAF.  The $gf$-values of the lines were taken from \citet{lambert96} and
from a compilation by R.E. Luck (1993, private communication).  We 
adopted NEXTGEN model atmospheres for low-mass stars
computed by \citet{nextgen99} interpolating within the grid when 
necessary.  We made use of the LTE stellar line analysis program
MOOG \citep{moog} to calculate the abundance of each Fe line based
on the measured equivalent width.  Effective temperatures 
(\teff) were set from the requirement
that the abundances of individual Fe lines be independent of lower excitation
potential.  The microturbulence was set by insisting that the abundances
of individual Fe lines show no trend against equivalent width.  By forcing agreement
between the Fe abundance derived from neutral and ionized lines, the gravity was
fixed.  This process required iteration until a consistent set of parameters
was obtained from which the Fe abundance was determined from the mean of
all Fe lines.  Estimated uncertainties in the model parameters are
$\delta$\teff=150K, $\delta$log g=0.3 dex, 
$\delta\xi_t$=0.3 km s$^{-1}$, and $\delta$[Fe/H]=0.2 dex.
We comment later upon the influence of these uncertainties upon the derived isotopic
ratios.  

\begin{figure}
\epsscale{1.0}
\plotone{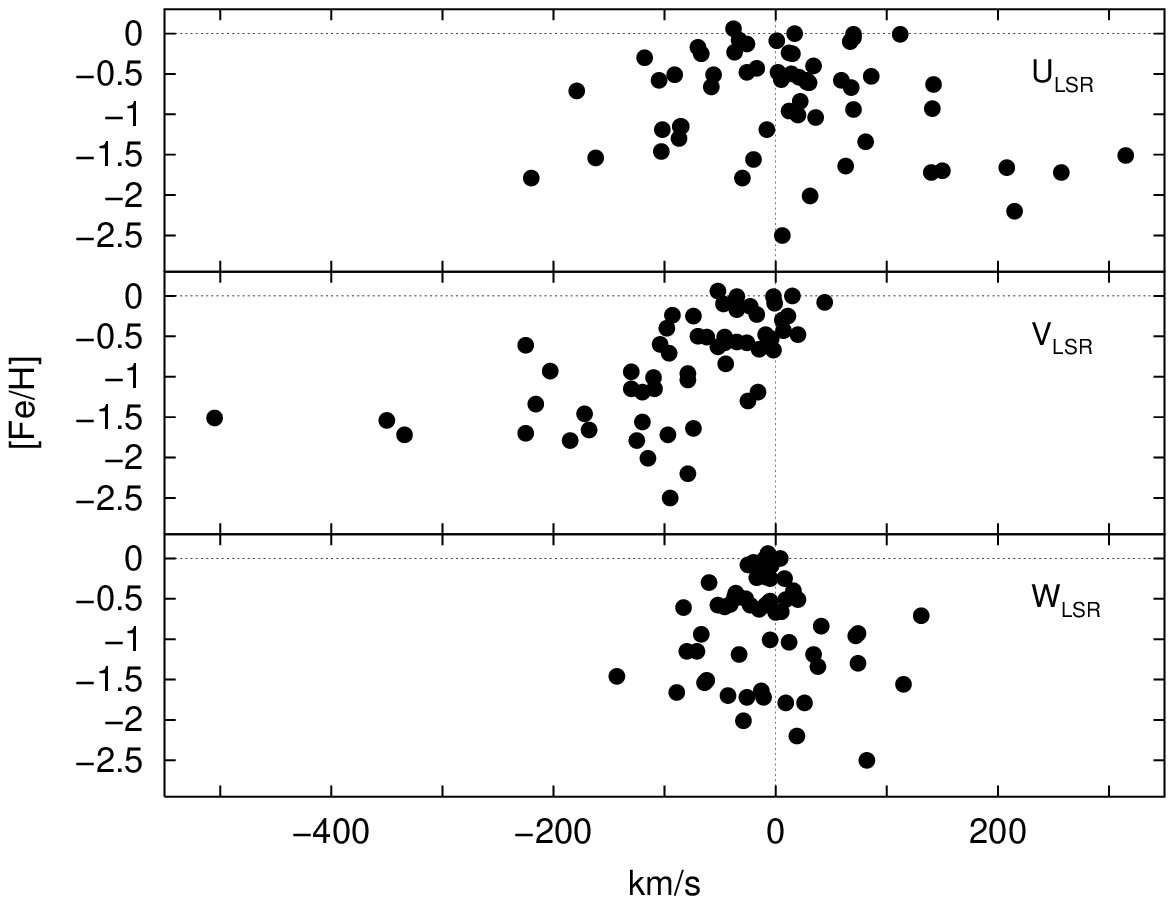}
\caption{Galactic space-velocity U, V, and W versus [Fe/H], where
U, V, and W are relative to the LSR.  As expected for a sample selected
against the thin disk, a considerable
number of stars lag the LSR (V$<-50$km s$^{-1}$).\label{fig:uvw}}
\end{figure}

The Galactic space-velocities of the sample were also derived in Paper I.  In
Figure \ref{fig:uvw} we plot U (positive towards the Galactic center), 
V (positive in the direction of Galactic rotation), and 
W (positive towards the north Galactic pole) corrected for the solar motion
with respect to the local standard of rest (LSR) versus [Fe/H].  
For the solar motion with respect to the LSR,
we adopted the \citet{lsr} values (+10,+5,+7) km s$^{-1}$ in (U,V,W).
An important
point that will arise later is that in order to find 
metal-poor stars, our targets were selected due to their
large reduced proper-motions and a majority are 
kinematically distinct from the thin disk. 

A subset of stars listed in Table \ref{tab:param} was not analysed in Paper I.
Seven stars, 5 giants and 2 dwarfs, were observed in 
August 2000 on the 2.7m Harlan J. Smith telescope
with a resolving power of $R = 120,000$ and a typical
S/N = 150 per pixel at 5140\AA.  A 20\AA~window 
around 5135\AA~was observed.  The stellar parameters for these stars
were taken from the literature as shown in Table \ref{tab:param}.  A 
further 7 dwarfs were observed with $R = 60,000$ and a typical
S/N = 200 per pixel at 5140\AA~in February 2003 on the
2.7m Harlan J. Smith telescope.  The stellar parameters for these stars
were derived in \citet{search2} using the method outlined in Paper I also 
described above.  These 7 stars were selected 
to have the kinematics of the thin disk with metallicities close to the solar value. 

\section{Analysis}
\label{sec:analysis}

\begin{figure}
\epsscale{1.0}
\plotone{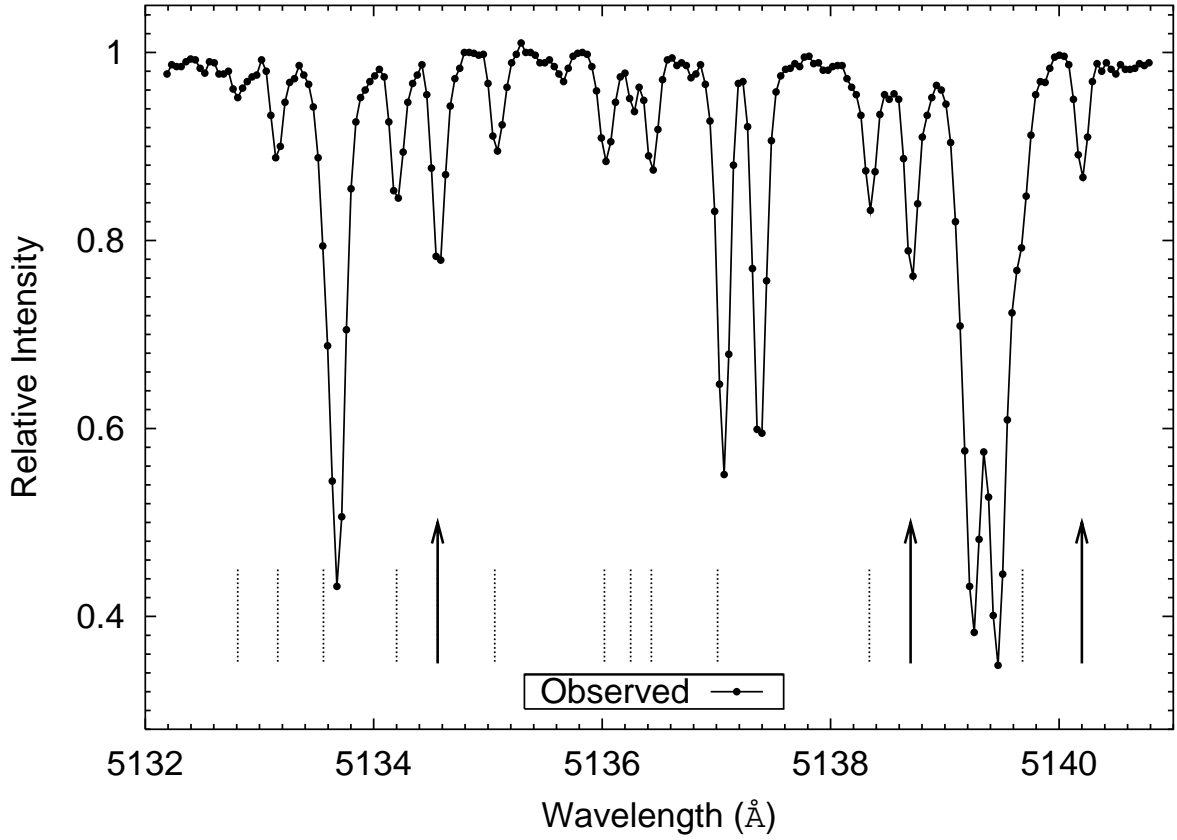}
\caption{Spectrum of G 17-25 from 5132 to 5141 \AA.  The positions of
various MgH A-X 0-0 and MgH A-X 1-1 lines are marked below the
spectrum.  The majority of MgH lines are unsuitable
for isotopic analysis.  The positions of the 3 features that we use to derive
the isotopic ratios are highlighted with arrows. \label{fig:features}}
\end{figure}

\begin{figure}
\epsscale{1.0}
\plotone{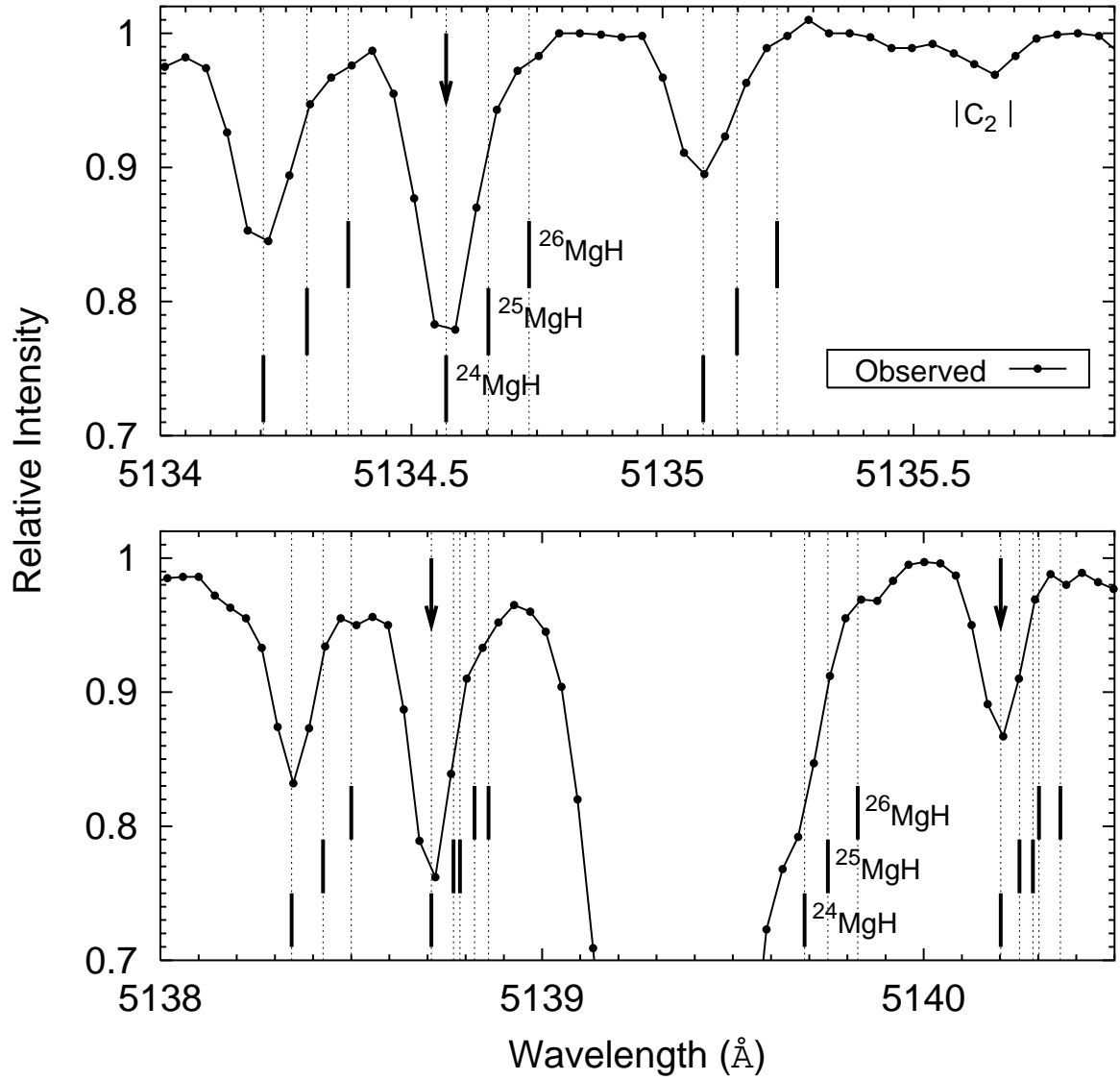}
\caption{Spectrum of G 17-25 from 5134 to 5136\AA~(upper) and
from 5138 to 5140.5\AA~(lower).  The positions of the 
$^{24}$MgH, $^{25}$MgH, and $^{26}$MgH lines are shown.  The lines 
used in the isotopic analysis to derive the ratios are marked by
arrows. \label{fig:features2}}
\end{figure}

The isotopic wavelength splitting in lines from the MgH A-X bands near 5140\AA~is small 
and so the $^{25}$MgH and $^{26}$MgH
lines are never fully resolved.  Instead, $^{25}$MgH and $^{26}$MgH contribute
a red asymmetry to the main $^{24}$MgH line.  Accordingly, synthetic spectra are
generated and fitted to the observed spectrum to derive the isotopic
ratio.  High resolving powers and high signal-to-noise ratios are essential
for measuring the Mg isotopic ratios.  
Our analysis techniques follow the method used by \citet{gl2000} and \citet{ml88}.  
Many MgH lines are present in the spectra of cool stars, though few are suitable for
isotopic abundance analysis due to blending by identified and unidentified
lines.  
Our Mg isotopic abundance ratios are derived from 3 MgH lines.
In Figure \ref{fig:features}, we show a region
of the spectrum that includes these 3 MgH lines which 
are a subset of the lines recommended
by \citet{ml88} for extraction of reliable Mg isotopic ratios.
These 3 features are identical to those used by \citet{gl2000} and 
are shown in more detail in Figure \ref{fig:features2}.
The feature at 5134.6\AA~is a blend of the
$Q_1(23)$ and $R_2(11)$ lines from the 0-0 band.  The red asymmetry
on the MgH features is due to the presence of $^{25}$MgH and $^{26}$MgH.  
The slightly weaker MgH features on either side of the 5134.6\AA~line
also exhibit red asymmetric wings though spectrum synthesis reveals that these lines
suffer from contamination and reliable isotopic ratios cannot be extracted 
from them \citep{tl80}.
The recommended feature at 5138.7\AA~is a blend of the 0-0 $Q_1(22)$
and 1-1 $Q_2(14)$ MgH lines.  The final recommended feature
at 5140.2\AA~is a blend of the 0-0 $R_1(10)$ and 1-1 $R_2(4)$
MgH lines.

\begin{figure}
\epsscale{1.0}
\plottwo{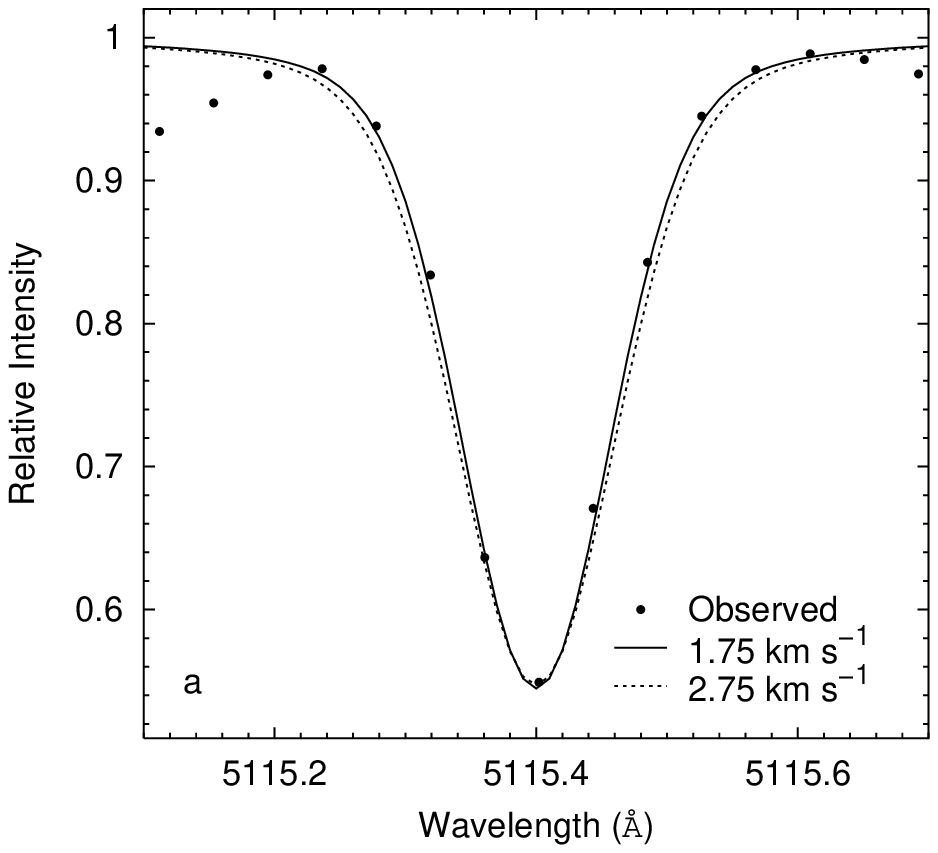}{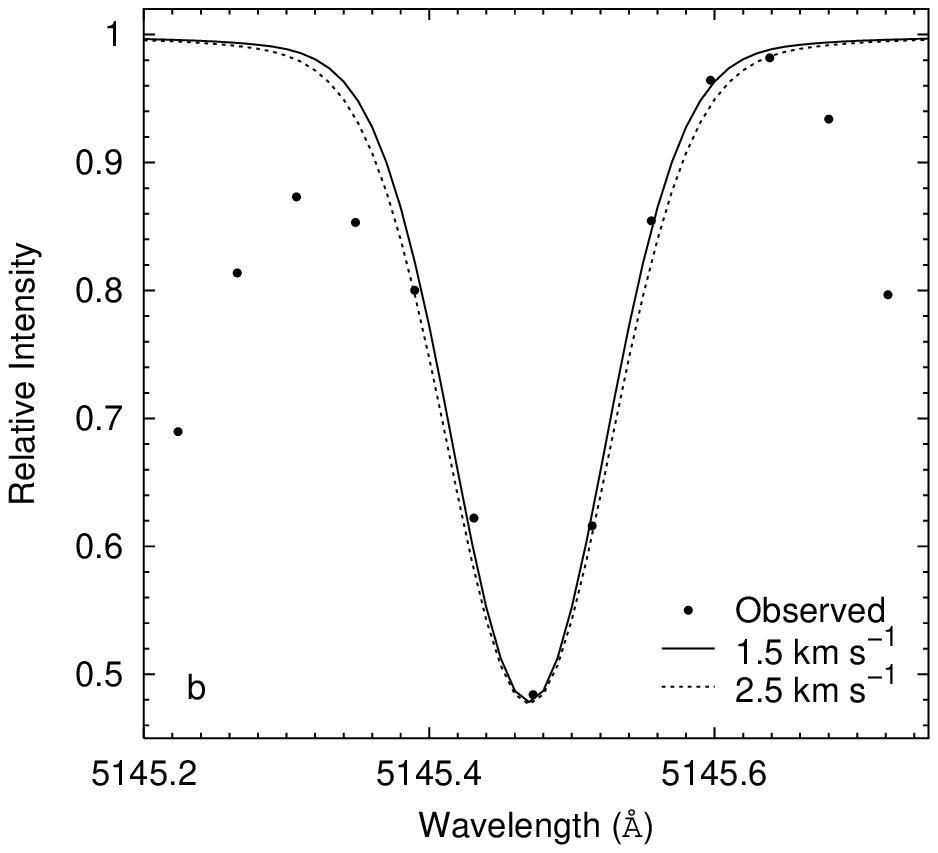}
\caption{The spectrum of G 9-13 showing the 
Ni\,{\sc i} at 5115.4\AA~and Ti\,{\scshape i} at 5145.5\AA~lines 
from which the macroturbulent broadening is determined.  
\label{fig:profile}}
\end{figure}

To determine the Mg isotopic abundance ratios, synthetic spectra
were produced using MOOG and fitted to the three MgH features.
Our list of atomic and molecular lines was identical
to the \citet{gl2000} list and included contributions from C, Mg, Sc, Ti, Cr,
Fe, Co, Ni, and Y.  The wavelengths of all
isotopic components were taken from \citet{ml88} and were based on direct measurements
of an MgH spectrum obtained using a Fourier transform spectrometer 
by \citet{bernath85}.  The instrumental profile was determined from Th lines 
in the spectrum of the Th-Ar comparison lamp.  The broadening due to 
macroturbulence was estimated by fitting the profiles of 
unblended lines of comparable depth to the MgH lines.
The chosen lines were Ni\,{\sc i} at 5115.4\AA~and Ti\,{\scshape i} 
at 5145.5\AA~where typical values for macroturbulence were 1.5--4.0 km s$^{-1}$ 
(see Figure \ref{fig:profile}).  These 2 lines gave the same 
macroturbulence within 0.25 km s$^{-1}$ and the larger value was adopted if 
there was a disagreement.  
Both the macroturbulent and instrumental broadening were assumed to
have a Gaussian form.  
We adjusted the Mg abundance to best fit the depths
of the MgH lines.  The $^{25}$Mg and $^{26}$Mg abundances were adjusted by
trial and error until the profile of a recommended feature was
fitted.  We did not require the abundances of $^{25}$Mg and $^{26}$Mg
to be equal.  For a given star, the final isotopic ratio of $^{24}$Mg:$^{25}$Mg:$^{26}$Mg 
was the value which provided the best fit to all 3 recommended features.
The best fit was determined by eye and the differences 
between the observed spectra and the best-fitting
syntheses were similar for all stars.  The derived Mg isotopic ratios are presented 
in Table \ref{tab:param}.  

\begin{figure}
\epsscale{1.0}
\plotone{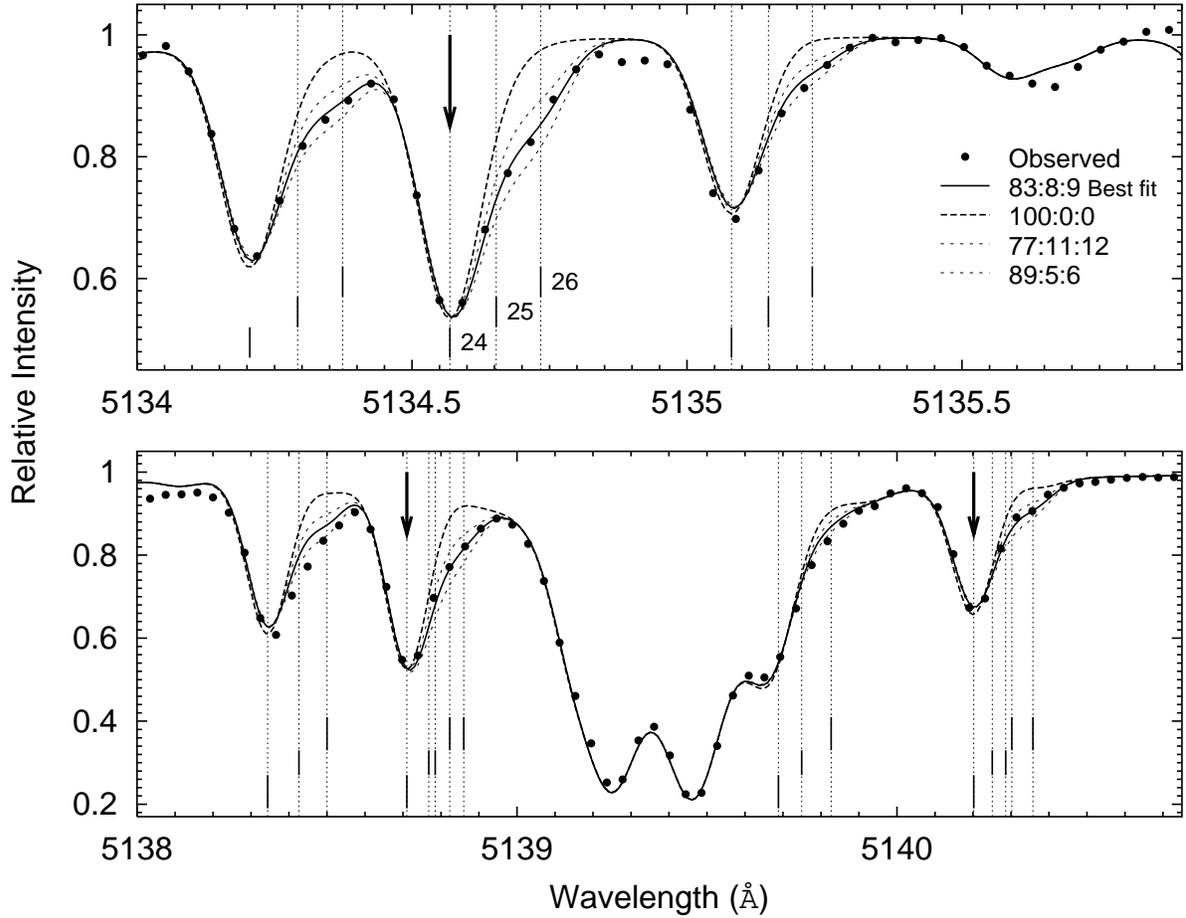}
\caption{The spectrum of G 9-13 from 5134 to 5136\AA~(upper)
and 5138 to 5140.5\AA~(lower).  The features we are interested in fitting
are marked from above by arrows.  The positions 
of the $^{24}$MgH, $^{25}$MgH, and $^{26}$MgH
lines are indicated by vertical dotted lines.  The closed circles represent the
observed spectrum.  The best fit to the 
recommended features is shown as a solid line for
$^{24}$Mg:$^{25}$Mg:$^{26}$Mg = 83:8:9.  The pure $^{24}$Mg synthesis
(100:0:0) is plotted as a dashed line and clearly provides a poor
fit.  The dotted lines represent
unsatisfactory ratios 77:11:12 and 89:5:6.  \label{fig:g9-13}}
\end{figure}

In Figure \ref{fig:g9-13}, we compare the observed and synthetic spectra
for G 9-13, a subdwarf with [Fe/H]=$-$0.58.  The strength of the MgH
features is comparable to the strength of the lines from which we
derived the macroturbulence (compare Figures \ref{fig:profile} and \ref{fig:g9-13}).
The red asymmetry of the MgH lines demands positive contributions 
from the $^{25}$Mg and $^{26}$Mg isotopes.
The Mg isotopic ratio $^{24}$Mg:$^{25}$Mg:$^{26}$Mg = 83:8:9 provides an
excellent fit to the three recommended MgH features.  
Note the poor fit of the pure $^{24}$Mg mix 
($^{24}$Mg:$^{25}$Mg:$^{26}$Mg = 100:0:0) 
to all the MgH lines.
Unsatisfactory ratios 77:11:12 and 89:5:6 are overplotted to give an
indication of the measurement uncertainties.

\begin{figure}
\epsscale{1.0}
\plotone{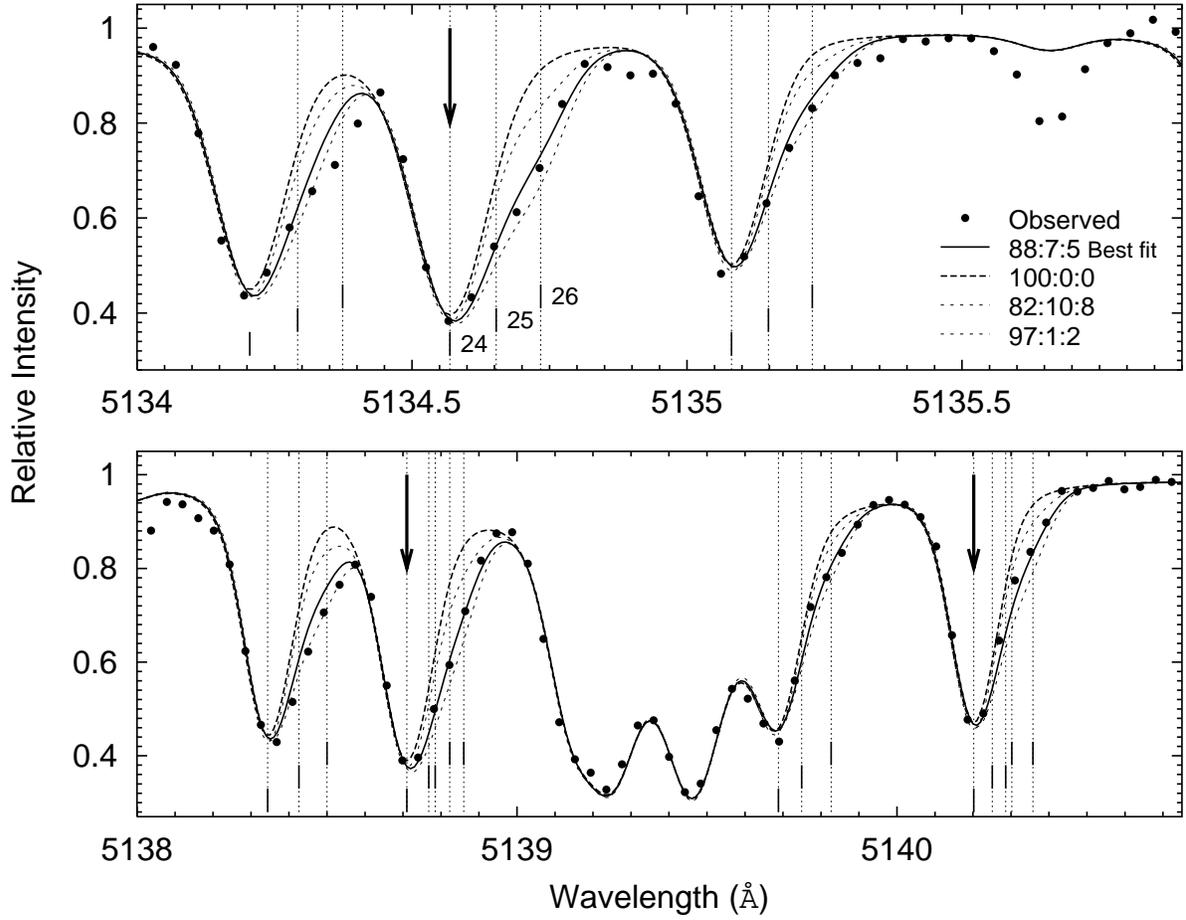}
\caption{The spectrum of LHS 2715 from 5134 to 5136\AA~(upper)
and 5138 to 5140.5\AA~(lower).  The best fit to the recommended
features shown as a solid line for $^{24}$Mg:$^{25}$Mg:$^{26}$Mg = 88:7:5.
The pure $^{24}$Mg synthesis
(100:0:0) is plotted as a dashed line and clearly provides a poor
fit.  The dotted lines represent
unsatisfactory ratios 82:10:8 and 97:1:2.  \label{fig:lhs2715}}
\end{figure}

In Figure \ref{fig:lhs2715}, we show the comparison of the observed and
synthetic spectra for LHS 2715, a subdwarf with [Fe/H]=$-1.56$.  The
MgH lines are very strong in this star.  The
best fitting Mg isotopic ratio is 88:7:5.
This ratio provides an excellent fit to all three recommended features.
Note again the poor fit of the pure $^{24}$Mg mix 
to all the MgH lines.
Unsatisfactory ratios 82:10:8 and 97:1:2 are overplotted to give an
indication of the measurement uncertainties.  Other MgH lines adjacent
to the recommended features show stronger red asymmetries indicating
the presence of unidentified blends.

\begin{figure}
\epsscale{1.0}
\plotone{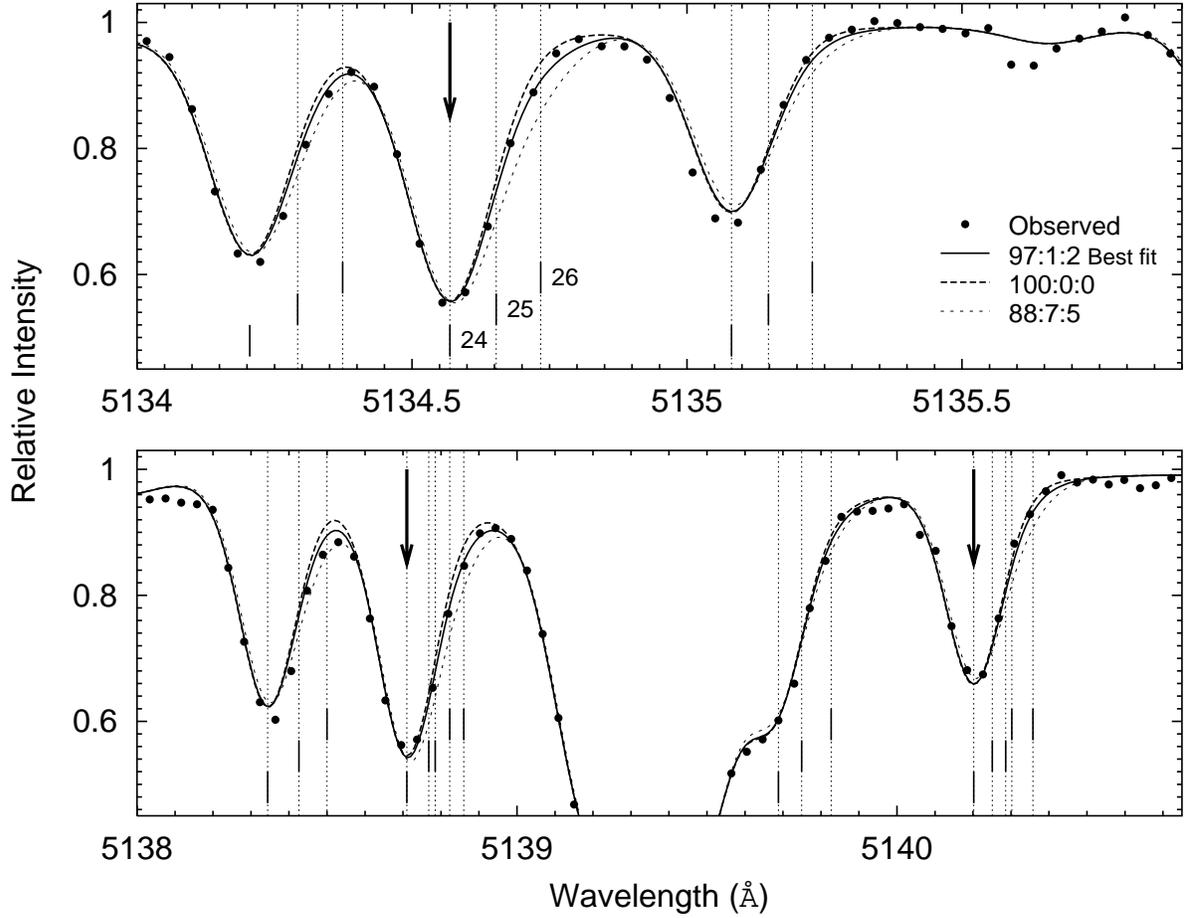}
\caption{The spectrum of PLX 5805 from 5134 to 5136\AA~(upper)
and 5138 to 5140.5\AA~(lower).  
The best fit to the 
recommended features is shown as a solid line for
$^{24}$Mg:$^{25}$Mg:$^{26}$Mg = 97:1:2.  The pure $^{24}$Mg synthesis
(100:0:0) is plotted as a dashed line.  The dotted line represent 
88:7:5 which is clearly a poor fit to the data.  \label{fig:plx5805}}
\end{figure}

In Figure \ref{fig:plx5805}, we plot observed and synthetic spectra for
PLX 5805, a subdwarf with [Fe/H]=$-1.72$.  This star also has strong
MgH lines, though not quite as strong as those seen in LHS 2715.  This difference 
is likely due to the \teff~difference 
rather than the difference in metallicity.  The best fitting isotopic ratio
is 97:1:2 which again provides a good fit to all 3 recommended features.
Within the uncertainties, the pure $^{24}$Mg synthesis 
also provides a reasonable fit to the data.  An unsatisfactory ratio 
88:7:5, the isotopic mix that fits LHS 2715, is overplotted to highlight 
the measurement uncertainties.  (This star was observed at R=35,000.)

When fitting synthetic spectra to observed line profiles, sources
of error include continuum fitting, microturbulence, macroturbulence,
identified and unidentified blends.  To understand the effects of 
these errors, we generated a synthetic spectrum assuming \teff=4300K,
log g=4.5, [Fe/H]=$-$2.5, $\xi_t$=0.5 km s$^{-1}$, macroturbulence=2.0 km s$^{-1}$,
$^{24}$Mg:$^{25}$Mg:$^{26}$Mg = 94:3:3, and R=60,000.  We added
noise to the synthesis to produce a spectrum with S/N=100 per pixel.  We then
treated this artificial spectrum as real data.  Assuming the correct
input parameters (\teff, log g, $\xi_t$, macroturbulence, [Fe/H]) our best
fitting ratio to the artificial spectrum was 96:2:2.  Incorrect choices 
for \teff, gravity or metallicity equally affect the
$^{24}$MgH, $^{25}$MgH, and $^{26}$MgH lines and so the measured isotopic ratios
are quite insensitive to the adopted model parameters.  
For strong lines, the microturbulence  
would affect the derived isotopic ratio as the strong $^{24}$MgH line is more 
sensitive to the adopted microturbulence compared 
to the weaker $^{25}$MgH and $^{26}$MgH 
lines.  We increased the microturbulence by 0.5 km s$^{-1}$ and measured 
97:1:2.  We decreased the continuum by 0.5\% and measured 96:2:2.  We 
increased the macroturbulence by 0.5 km s$^{-1}$ and measured 97:2:1.
An identical test was performed on a synthetic spectrum generated assuming \teff=4800K,
log g=4.5, [Fe/H]=$-$0.25, $\xi_t$=0.6 km s$^{-1}$, macroturbulence=2.5 km s$^{-1}$,
$^{24}$Mg:$^{25}$Mg:$^{26}$Mg = 80:10:10, R=60,000, and S/N=90 per pixel.
In both tests we found that errors in the model parameters (\teff, log g, $\xi_t$,
macroturbulence, continuum) are about b$\pm$1 and
c$\pm$1 when expressing the ratios as $^{24}$Mg:$^{25}$Mg:$^{26}$Mg =
(100-b-c):b:c.
Inspection of Figures \ref{fig:g9-13}, \ref{fig:lhs2715}, 
and \ref{fig:plx5805} suggests that the
uncertainties in determining the best fit are at the level b$\pm$3 and
c$\pm$3 when expressing the ratios as $^{24}$Mg:$^{25}$Mg:$^{26}$Mg =
(100-b-c):b:c.  The ratio $^{25}$Mg/$^{24}$Mg is less accurately
determined than the $^{26}$Mg/$^{24}$Mg ratio due to the larger isotopic shift
of $^{26}$MgH.  That is, $^{26}$MgH is less blended with the strong $^{24}$MgH line 
compared to $^{25}$MgH.  

There are 6 stars in common with the \citet{gl2000} sample, almost a third
of their sample.  In Table \ref{tab:comp}, we compare the Mg isotopic ratios
derived in the two different studies.  Since the analysis techniques are
essentially identical, the differences between the two studies can be
attributed mainly to the quality of the data.  The Gay \& Lambert data
are superior in both resolution ($R = 150,000$ versus $R = 60,000$) and S/N
(150 versus 90 per pixel).  The agreement is excellent
for the 6 stars common to both samples, particularly for the ratio
$^{26}$Mg/$^{24}$Mg.  Gay \& Lambert state that their errors are around
b$\pm$2 and c$\pm$2 when expressing the ratios as $^{24}$Mg:$^{25}$Mg:$^{26}$Mg =
(100-b-c):b:c.  That is, our estimated errors are 50\% larger than those
reported by Gay \& Lambert which reflects the difference in data quality.

In Table \ref{tab:comp2}, we compare 5 Mg isotopic ratios measured in giants with
those reported by \citet{shetrone96b}.  Shetrone's data were taken at
$R = 60,000$ with S/N ranging from about 80 to 160 per resolution
element.  Our data were taken at $R = 120,000$ with S/N around 150
per pixel (210 per resolution element).  More importantly, the Shetrone study employed a
different set of MgH lines to derive the isotopic ratios.  Although Shetrone could not
distinguish the contribution of $^{25}$Mg from $^{26}$Mg, we find a fair agreement for 3 stars.
For the other 2 stars we find a poor agreement with Shetrone's values 
due to the different data quality and choice of MgH lines.

\section{Discussion}
\label{sec:discussion}

\subsection{Observed trends}

\begin{figure}
\epsscale{1.0}
\plotone{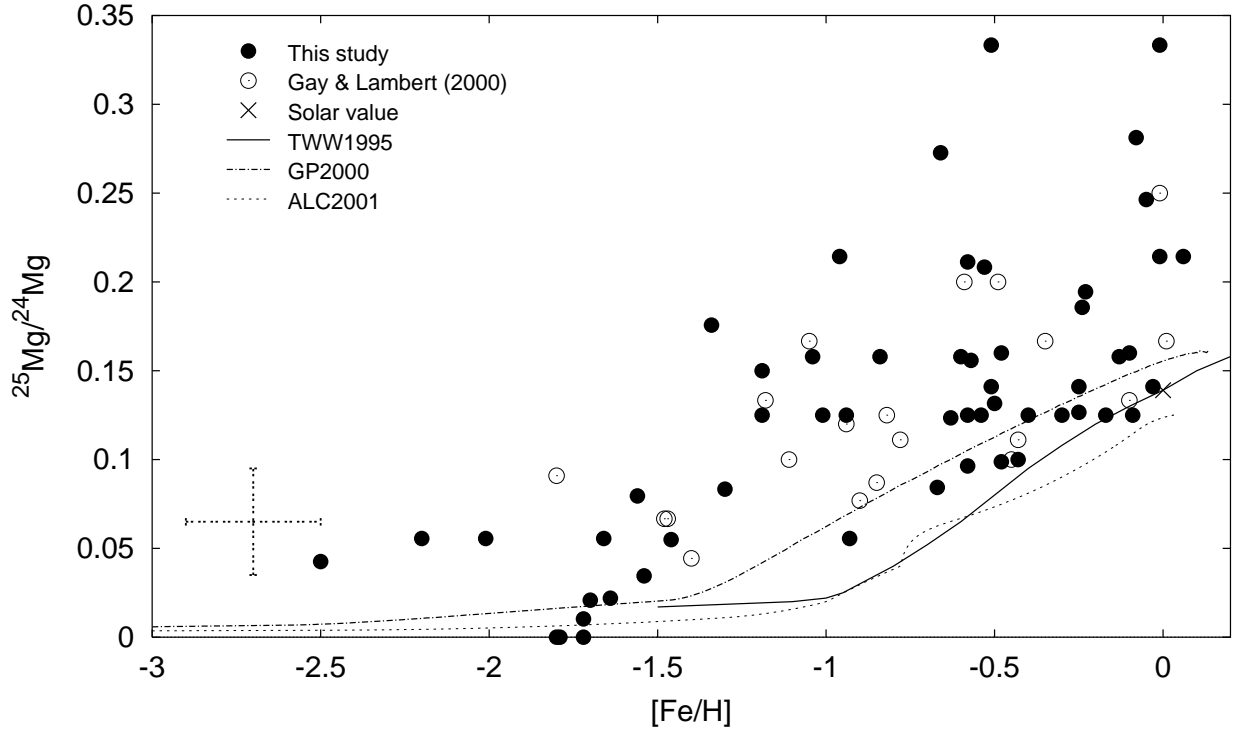}
\caption{Evolution of isotopic ratio $^{25}$Mg/$^{24}$Mg versus
[Fe/H].  The filled circles represent data from this study, the open
circles represent the \citet{gl2000} data, and the cross marks
the solar value.  (The Gay \& Lambert value is plotted for the stars 
in common.)  A representative error bar is given.  The three
lines are the \citet{timmes95}, \citet{goswami00}, and \citet{alc01}
predictions. \label{fig:25mg}}
\end{figure}

\begin{figure}
\epsscale{1.0}
\plotone{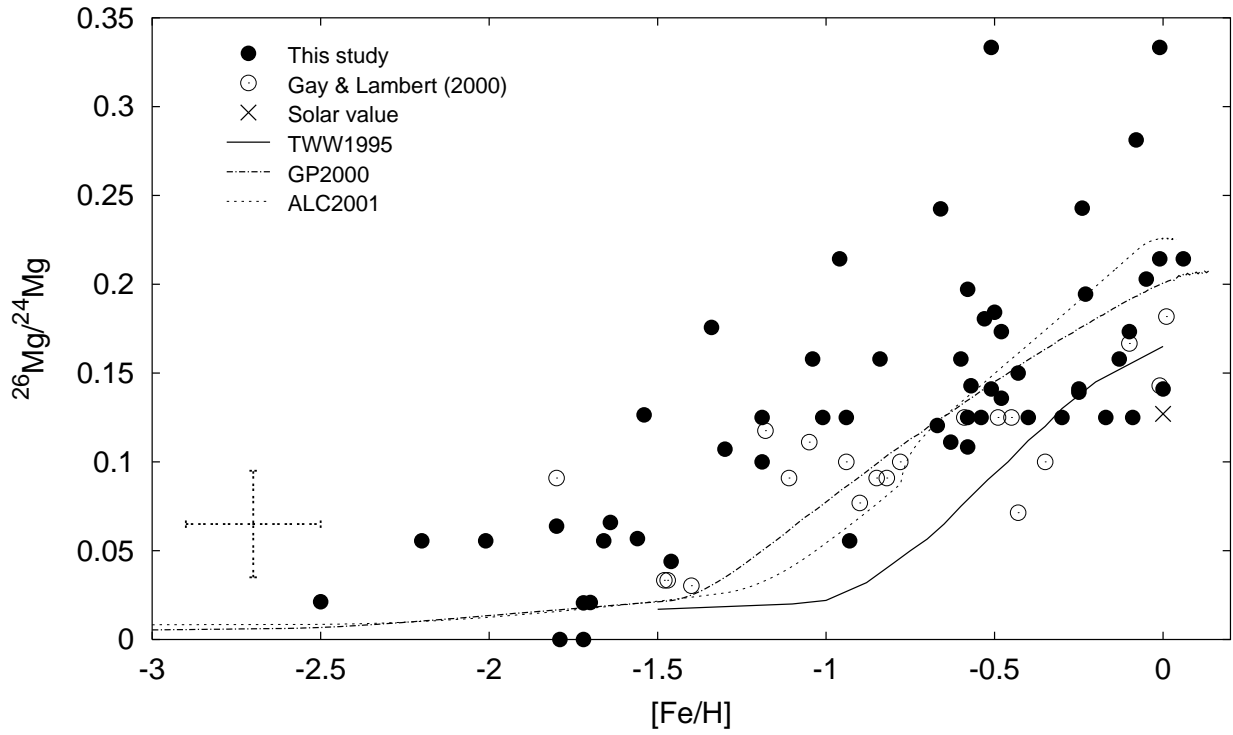}
\caption{Same as Fig.~\ref{fig:25mg} but for $^{26}$Mg/$^{24}$Mg.
\label{fig:26mg}}
\end{figure}

\begin{figure}
\epsscale{1.0}
\plotone{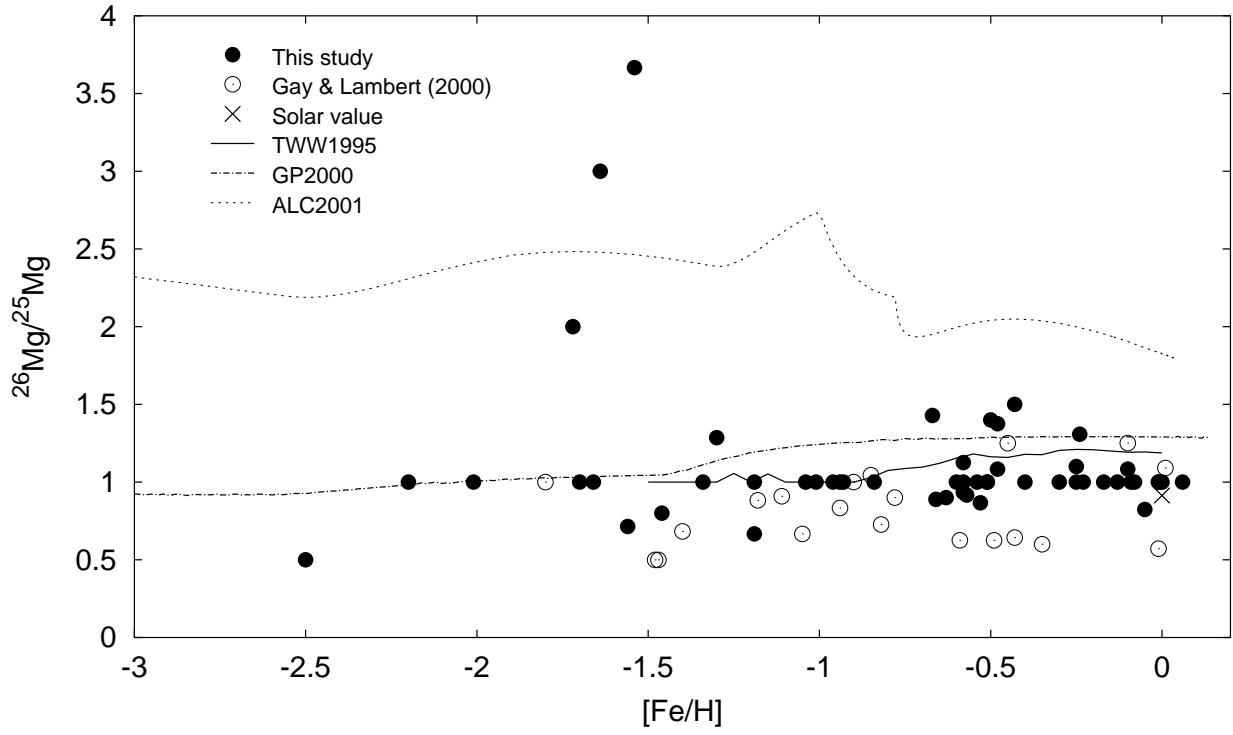}
\caption{Same as Fig.~\ref{fig:25mg} but for $^{26}$Mg/$^{25}$Mg.
\label{fig:2526mg}}
\end{figure}

The evolution of the Mg isotopic ratios with metallicity is shown in
Figures \ref{fig:25mg}, \ref{fig:26mg}, and \ref{fig:2526mg}.  The 
ratios $^{25}$Mg/$^{24}$Mg, $^{26}$Mg/$^{24}$Mg, and $^{26}$Mg/$^{25}$Mg are plotted
against [Fe/H].   We combine our sample with data from 
the \citet{gl2000} study where the total sample consists of 75 different stars.
The observed ratios $^{25}$Mg/$^{24}$Mg and $^{26}$Mg/$^{24}$Mg decline
with decreasing metallicity.  Below [Fe/H]=$-$1.5, there are 6 stars that
show non-zero ratios $^{26}$Mg/$^{24}$Mg$>$0.05.  Even taking into account the errors,
it is unlikely that these 6 stars have $^{26}$Mg/$^{24}$Mg=0.
Around solar metallicity, the solar ratios
$^{25}$Mg/$^{24}$Mg and $^{26}$Mg/$^{24}$Mg lie at the lower boundary.
The ratio $^{26}$Mg/$^{25}$Mg is essentially
constant and centered at unity.  The solar ratio, $^{26}$Mg/$^{25}$Mg,
does not appear to be unusual.  In Figure \ref{fig:2526mg}, there are 
3 stars with large
excesses of $^{26}$Mg, or equivalently, underabundances of $^{25}$Mg.
These 3 stars have ratios of 87:3:11, 91:2:6, and 97:1:2 and within
the measurement uncertainties ($^{25,26}$Mg $\pm$ 3), all stars could
have $^{25}$Mg $\simeq$ $^{26}$Mg.

\begin{figure}
\epsscale{1.0}
\plotone{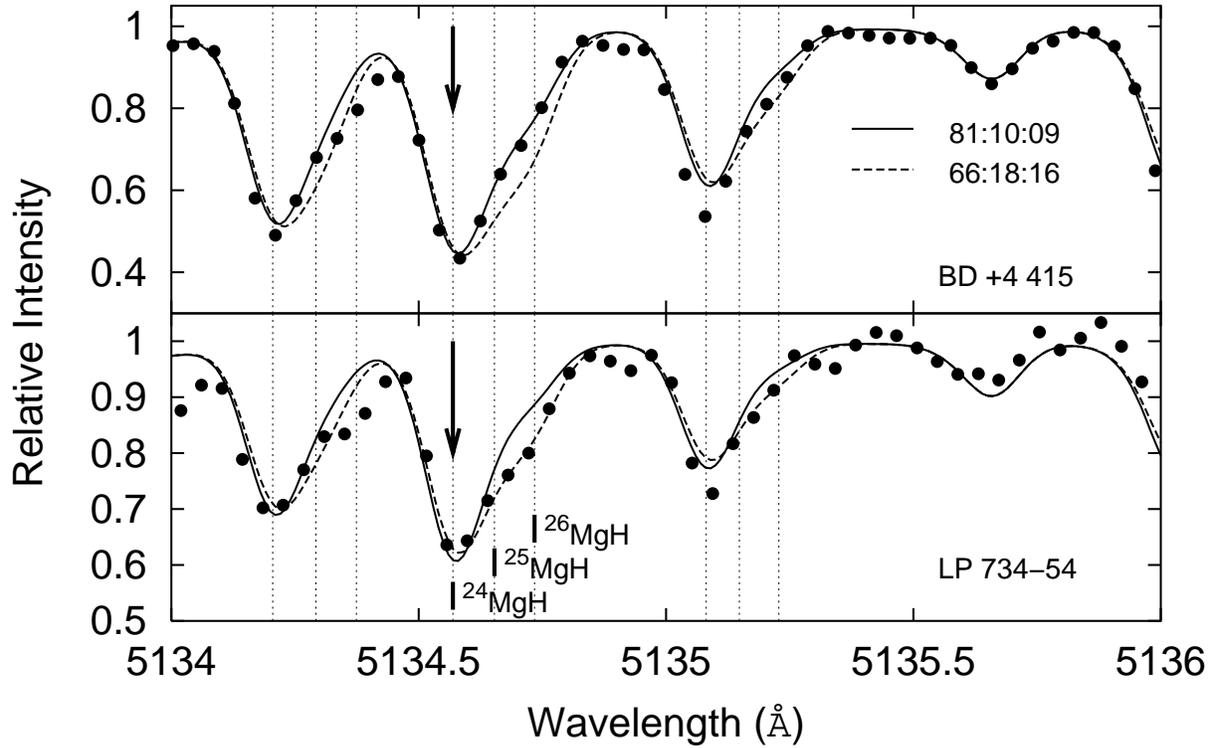}
\caption{Spectra of BD +4 415 (upper) and LP 734-54 (lower).
In both panels, the observed spectra are 
shown as closed circles and two different
isotopic ratios are plotted, 81:10:9 (solid line) and 68:18:16
(dashed line).  The ratio 81:10:9 is the best fit
for BD +4 415 while 68:18:16 is the best fit for LP 734-54.
The positions of the $^{24}$MgH, $^{25}$MgH, and $^{26}$MgH
lines are indicated.  The red asymmetries are 
very different between these two stars and
the difference in the Mg isotopic ratios exceeds the measurement errors.
\label{fig:scatter}}
\end{figure}

Gay \& Lambert showed that at a given [Fe/H], there was 
a scatter in the isotopic ratios exceeding the measurement errors.
For our sample, we confirm the spread in $^{25,26}$Mg/$^{24}$Mg and
the dispersion appears to increase with increasing metallicity.
Figures \ref{fig:25mg} and \ref{fig:26mg} show a real scatter at low
metallicity.  Earlier we plotted observed and synthetic spectra 
for 2 stars with similar metallicities and atmospheric parameters 
LHS 2715 ([Fe/H]=$-1.56$) and PLX 5805 ([Fe/H]=$-1.72$).  
The best fitting ratio to LHS 2715 is 88:7:5 and the best
fitting ratio to PLX 5805 is 97:1:2.  In Figure \ref{fig:lhs2715},
we overplot the ratio 97:1:2 upon the spectrum of LHS 2715 to show
that it provides a poor fit.  Likewise in Figure \ref{fig:plx5805},
we overplot the ratio 88:7:5 upon the spectrum of PLX 5805 and find
that it provides a poor fit.  At higher metallicity, Figure \ref{fig:scatter}
shows that the scatter certainly exceeds the measurement errors.  
In this figure, we plot the observed spectra of 2 stars with similar metallicities, 
BD +4 415 ([Fe/H]=$-0.63$) and LP 734-54 ([Fe/H]=$-0.66$) and almost identical
atmospheric parameters.
The measured isotopic ratios are 81:10:9 for BD +4 415 and 66:18:16 for
LP 734-54.  The best ratio for BD +4 415 provides a very poor fit when compared with
the observed spectrum of LP 734-54.  Similarly, the isotopic ratio
offering the best fit to LP 734-54 provides a poor fit to the
observed spectrum of BD +4 415.  In short, there is a real scatter
in the measured Mg isotope ratios at low and high metallicities.

For a given metallicity, our stars generally exhibit higher
isotopic ratios ($^{25}$Mg/$^{24}$Mg and $^{26}$Mg/$^{24}$Mg)
than those within the Gay \& Lambert sample.  This is particularly
true at the higher metallicities, [Fe/H]$>-1.0$.  As mentioned
earlier, Table \ref{tab:comp} shows the Mg isotopic ratios for 
6 stars common to both studies where the agreement is excellent.  This 
agreement suggests that the high values of $^{25}$Mg/$^{24}$Mg and 
$^{26}$Mg/$^{24}$Mg are real.  

BD +30 4633 and LP 790-19 both show large ratios $^{25,26}$Mg/$^{24}$Mg $\simeq$ 0.3.  
The MgH lines in these stars are rather strong.  Since these unusually large isotopic
ratios were found in stars with strong MgH lines, we conducted an additional test to
determine if there were systematic errors
in the analysis.  We changed the macroturbulence and microturbulence by 0.5 km s$^{-1}$.  
Assuming these new values, we remeasured the Mg isotope ratios.  
In both cases, the measured Mg isotope 
ratio changed by only b$\pm 2$ and c$\pm 2$ when expressing the ratio as 
$^{24}$Mg:$^{25}$Mg:$^{26}$Mg = (100-b-c):b:c.  As discussed earlier, 
the analysis techniques provide firm limits on the error bars.  
Giants in metal-poor globular clusters have shown 
$^{25+26}$Mg/$^{24}$Mg $\simeq$ 1.0 \citep{shetrone96b} and 
$^{24}$Mg:$^{25}$Mg:$^{26}$Mg = 53:9:39 \citep{6752},
values comparable to and greater than the highest isotopic ratios 
found in this study.  

\begin{figure}
\epsscale{0.8}
\plotone{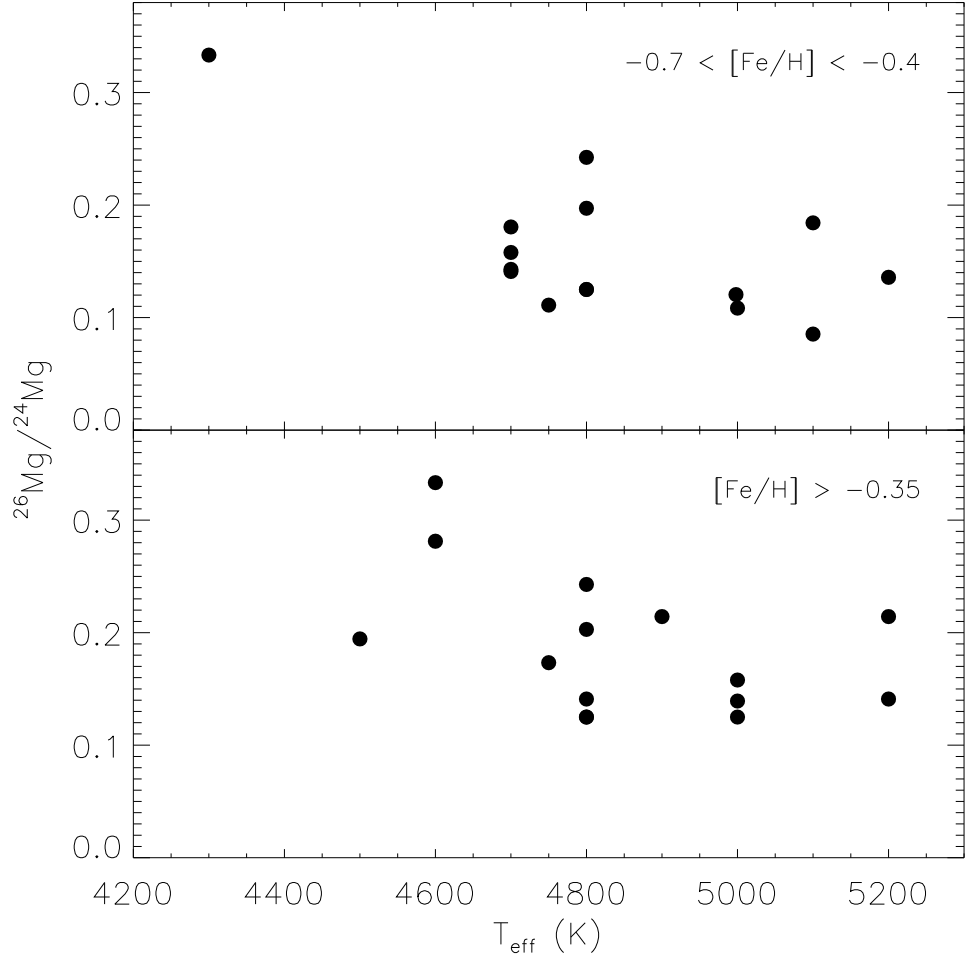}
\caption{Mg isotope ratio $^{26}$Mg/$^{24}$Mg versus \teff~for
two [Fe/H] intervals.  In the upper panel, the data cover 
$-0.7 <$ [Fe/H] $< -0.4$ and in the lower panel, the data cover 
[Fe/H] $> -0.35$. The ratio $^{26}$Mg/$^{24}$Mg may increase with decreasing \teff.
\label{fig:iso.teff}}
\end{figure}

While there is no obvious trend of $^{25,26}$Mg/$^{24}$Mg with \teff, it is curious
that the two stars with $^{25,26}$Mg/$^{24}$Mg $\simeq$ 0.3 are both cool and
have strong MgH lines.  In Figure \ref{fig:iso.teff}, we show the
$^{26}$Mg/$^{24}$Mg ratio for dwarfs versus \teff~for two [Fe/H] intervals.  There
is a hint that the isotopic ratio increases with decreasing \teff.  If confirmed by further 
exploration of cool dwarfs, it may point to an inadequacy of the classical model 
atmospheres and/or the assumption of LTE for MgH line formation.

An assumption of classical atmospheres is that of homogeneous
layers.  Suppose the real atmosphere consists of cool and hot columns
with MgH lines strongly represented in the spectrum of the
former but more weakly in the latter.  The continuum from the hot columns
dilutes the MgH lines from the cool columns.  In analysing the 
combined spectrum of the cool and hot columns with a classical
atmosphere, one underestimates the level of saturation of the
MgH lines and overestimates the $^{25}$Mg/$^{24}$Mg and
$^{26}$Mg/$^{24}$Mg ratio for those stars in which the MgH lines
are strong.  Just such an effect was suggested by \citet{lambert71}
to account for a report of high (non-terrestrial)
isotopic ratios from spectra of sunspot umbrae; the bright
umbral dots seen in high-resolution images of sunspots serve as the
hot columns in this example.  Velocity differences between cool and
hot columns and within a column would be an additional factor
not included in our analysis, but the isotopic shifts are several
times the expected Doppler shifts arising from stellar granulation.
A thorough analysis of MgH (and other) lines from the weakest to the strongest
might shed light on the extent of the inhomogeneities. 
It should be noted 
that stars of very similar atmospheric parameters may show quite different
isotopic ratios suggesting that, if a failure of classical atmospheres is
the responsible factor, that the inhomogeneities are not simply
dependent on effective temperature, surface gravity, and metallicity.

Departures from LTE could lead to systematic errors in the isotopic
ratios.  Suppose that the line source function exceeds everywhere
the local Planck function.  Then, the ratio of the $^{25}$MgH and $^{26}$MgH
line to the $^{24}$MgH line will be less than expected in LTE.
An LTE analysis of these lines will likely result in a systematic
overestimate of the isotopic ratios.  The fact that the scatter in
isotopic ratios among the stars is present for almost identical
stars, independent of effective
temperature and present in dwarfs and giants suggests that
departures from LTE are not a major influence on the derived
ratios. 

\subsection{Model predictions}

Three different models predict the evolution of the
elements from carbon through zinc, including the 
Mg isotopes: \citet{timmes95} (hereafter TWW1995), 
\citet{goswami00} (hereafter GP2000), and \citet{alc01} (hereafter ALC2001). 
Underpinning each of these models are assumptions
regarding the dynamical evolution of the Galaxy, the initial mass function,
the star formation rate, and the stellar yields.  
We will briefly compare and
contrast various assumptions between the three different models with an emphasis on
those that would affect the predicted evolution of the 
Mg isotopic ratios.  Of the major ingredients, 
only the stellar yields can be calculated from first principles,
though particular reaction rates may be plagued by 
uncertainties.  For massive stars that die
as Type II supernovae, all three models rely upon the \citet{woosley95} yields
that include all isotopes of Mg along with Fe.  
TWW1995 also consider the contribution from intermediate- to low-mass stars that
become planetary nebulae and intermediate- to low-mass stars that become
Type Ia supernovae.  
TWW1995 assumed a Salpeter initial mass function 
and a dynamical model for the Galactic disk.  The GP2000 study differs from
TWW1995 by using ``appropriate models for both the halo and disk'' along with the
\citet{kroupa93} initial mass function ``which presumably describes the distribution 
of stellar masses better than the Salpeter IMF''.  GP2000 deliberately neglect the contribution
from intermediate mass stars in order to gauge ``to what extent those stars
(or other sources) are required to account for the observations.''
ALC2001 also treat the halo and disk
independently and use ``metallicity-dependent stellar yields for the whole range
of stellar masses considered''.  ALC2001 also adopt the \citet{kroupa93} initial mass function.
For all three models, the Type II and Type Ia
supernovae are responsible for the production of Fe whereas the Type II supernovae
are responsible for the production of the Mg isotopes.  
The TWW1995 predictions offer a reasonable fit to observed elemental abundances 
(e.g., Mg, Si, and Ca) provided the Fe yields from Type II supernovae are reduced by a factor of 2.
The observed elemental abundances (e.g., Mg, Si, and Ca) are reproduced by both the GP2000 and ALC2001 
predictions.

In Figures \ref{fig:25mg}, \ref{fig:26mg}, and \ref{fig:2526mg} we overplot the TWW1995,
GP2000, and ALC2001 predictions along with our measured Mg isotopic ratios
and the \citet{gl2000} measurements.  The three predictions are qualitatively similar
which is unsurprising as all three models rely upon the same source for yields
from massive stars. 
The abundances of $^{25}$Mg and $^{26}$Mg with respect to $^{24}$Mg fall away
with decreasing metallicity.  This decrease in $^{25}$Mg/$^{24}$Mg and 
$^{26}$Mg/$^{24}$Mg is due to the decrease in the abundance of $^{22}$Ne,
the seed from which the heavy isotopes are produced via
$^{22}$Ne($\alpha,n)^{25}$Mg($n,\gamma)^{26}$Mg.  The predicted 
Mg isotopic ratios suggest that once the metallicity in massive stars reaches a
critical value, the production of $^{25}$Mg and $^{26}$Mg becomes
increasingly efficient.  This is consistent with the Woosley \& Weaver massive
star yields.  At low metallicities, the non-zero plateau in the
predicted isotopic ratios reflects the primary production of the neutron-rich 
Mg isotopes.  Primary production takes place because massive 
metal-poor stars produce C, N, and O which can then
be burned into $^{22}$Ne.
The abundances of $^{26}$Mg and $^{25}$Mg are predicted to be
equal by TWW1995 and GP2001, but ALC2001 predict a higher value of $^{26}$Mg/$^{25}$Mg.

Our measured Mg isotopic ratios $^{25}$Mg/$^{24}$Mg and $^{26}$Mg/$^{24}$Mg 
are consistently higher than the predictions.  The discrepancy is unlikely to stem
from incorrect values for [Fe/H] as errors exceeding 0.6 dex would be required and in
Paper I we showed that our derived metallicities agreed with published values.
One possibility for accounting for the low predictions is that the
yields from massive stars may underestimate the production of $^{25}$Mg and 
$^{26}$Mg.  In our current sample there is evidence that below [Fe/H]=$-$1.5
the Mg isotope ratios are higher than the predicted plateau.  
Observations of more Mg isotopic ratios at low metallicities, around
the predicted plateau, would prove a powerful tool for testing theoretical yields 
from massive metal-poor stars.  
Another possibility for explaining the low predictions for $^{25,26}$Mg/$^{24}$Mg, 
one previously raised by Gay \& Lambert, is that 
there is an additional source of the minor Mg isotopes.  AGB stars can
produce the neutron-rich isotopes and their role will be discussed in the following 
section.  We stress that none of 
the models include the contribution to $^{25}$Mg and $^{26}$Mg provided by
the ejecta from intermediate mass AGB stars.  Therefore, the model predictions should be regarded
as a lower limit to the observed Mg isotopic ratios.  Indeed, these predictions provide
a good fit to the lower envelope.

We re-emphasize a point made by Gay \& Lambert regarding the predicted
Mg isotopic ratios.  The assumed mass cut for Type II supernovae affects the
amount of Fe that is ejected while the amount of lighter elements ejected,
including Mg, are unaffected by the position of the mass cut.  Adjustments to
the mass cut have the effect of translating the predicted value of 
$^{25,26}$Mg/$^{24}$Mg along the [Fe/H] axis.  Both GP2000 and ALC2001
take the TWW1995 suggestion that the Fe yields from Type II supernovae need to be
reduced by a factor of 2 in order to match observations.  This 
corresponds to a factor of 0.3 in the log and
translating the TWW1995 predicted curve to lower metallicity effectively 
superimposes all predictions shown in Figures \ref{fig:25mg} and \ref{fig:26mg}.
Likewise, the assumed ratio of Type II to Type Ia supernovae exerts control
over the shape of the predicted curve.  Type II supernovae synthesize and
eject the Mg isotopes and iron whereas Type Ia supernovae 
return iron to the interstellar medium (ISM).  Most nucleosynthetic predictions
show that Type Ia supernovae produce one or two orders of magnitude less Mg
than Type II supernovae.  Different initial mass functions offer distinct
ratios of Type II to Type Ia supernovae affecting both the onset of the
increase in the isotopic ratios as well as the slope of the predicted curve.
We note that such adjustments cannot explain the scatter in the isotopic ratios 
and would adversely affect the fits to the observed run of elemental abundances (e.g., Mg, Si, Ca)
with respect to Fe.

\subsection{Role of AGB stars}

None of the predictions from the Galactic chemical evolution models take into account 
the yields of Mg from AGB stars.  In all of the papers describing the models,
a discussion is included to acknowledge and recognize
that intermediate mass AGB stars can produce
$^{25}$Mg and $^{26}$Mg.  Sufficiently massive AGB stars have He-shells that can
reach temperatures ($\sim 300 \times 10^6 K$) at which the neutron source 
$^{22}$Ne($\alpha,n)^{25}$Mg is activated.  Hot bottom burning can take place
if the base of the convective envelope reaches the top of the H-shell.  If
hot bottom burning occurs at
temperatures exceeding $\sim 90 \times 10^6 K$, the Mg-Al chain can
deplete $^{24}$Mg to produce $^{25}$Mg and $^{26}$Mg.
\citet{karakas03} have shown that 6$M_\odot$ AGB stars with [Fe/H]=$-0.7$ produce an envelope 
with large amounts of $^{25}$Mg and $^{26}$Mg and essentially no $^{24}$Mg.
This envelope is then ejected as the
star becomes a planetary nebula enriching the ISM with the 
neutron-rich Mg isotopes.  In contrast to supernovae which eject material
at high velocities ($\sim$5,000 km s$^{-1}$), AGB stars eject gas
at low velocities ($\sim$10 km s$^{-1}$).
The ejecta from AGB stars may therefore be confined to a more localized region
and inhomogeneities in the ISM are possible until the ISM is mixed by 
supernovae.
The timescale for star formation relative to mixing in the ISM is then an
important factor.  Stars formed from a region of the interstellar medium recently
polluted by massive AGB stars will therefore be enriched in $^{25}$Mg, $^{26}$Mg,
and $s$-process elements.

\citet{gl2000} measured overabundances of $^{25}$Mg and $^{26}$Mg in a 
number of stars known to have higher than usual abundances
of $s$-process elements.  These elemental and 
isotopic compositions are consistent with the 
idea that the stars' natal clouds were enriched by ejecta
from intermediate-mass AGB stars.  
One star in our sample, BD +5 3640, has already been shown to be a CH star
\citep{tomkin99}.  We find that it has high ratios of $^{25}$Mg/$^{24}$Mg
and $^{26}$Mg/$^{24}$Mg compared to other stars of similar metallicity.  Mass 
transfer from a companion AGB star, now an unseen white dwarf,
changed BD +5 3640 into a CH star and produced the observed enhancements in
$^{25}$Mg, $^{26}$Mg, and $s$-process elements.

Our Mg isotopic ratios exceed the predictions
calculated under the assumption that only massive stars produce $^{25}$Mg and $^{26}$Mg.
Inclusion of the yields of $^{25}$Mg and $^{26}$Mg from AGB stars would increase the
predicted ratio of $^{25,26}$Mg/$^{24}$Mg and introduce a scatter that would
qualitatively match our measured ratios.  \citet{busso01}
have shown that 5M$_\odot$ AGB stars produce considerable amounts of 
Y and Zr (and other $s$-process elements) in
addition to the neutron-rich Mg isotopes.  Busso et al.\ did not include hot bottom 
burning in their models.  Comparison of the Busso et al.\ and Karakas \& Lattanzio
yields show that hot bottom burning is the dominant production site of the minor Mg
isotopes in 5M$_\odot$ AGB stars.  Quantitative yields from AGB stars for a variety
of masses and metallicities need to be incorporated into comprehensive models
of Galactic chemical evolution to provide thorough predictions of the
evolution of the Mg isotopes.  Simmerer et al.\ (2003, private communication)
are also investigating the role of AGB stars from a different perspective.  
From measurements of the ratio of La ($s$-process) to Eu ($r$-process), their data suggest
the contribution of AGB stars to Galactic chemical evolution commences around [Fe/H]
$\simeq -2.0$.  This evidence supports the results of previous studies based on the
abundances of Ba and Eu (e.g., \citealt{spite78} and \citealt{burris00}).

We performed a back-of-the-envelope calculation to investigate whether enhancements
of Zr (a representative light $s$-process element) should be detectable 
in stars with high ratios of $^{26}$Mg/$^{24}$Mg.  In this
simple exercise, we assumed that the enrichment of $^{26}$Mg is a consequence of
pollution by AGB ejecta.  Taking the Karakas \& Lattanzio Mg yields for a 5$M_\odot$ Z=0.004 AGB
star, we found that a mix of 200 parts ambient material to 1 part AGB ejecta was required to
produce $^{26}$Mg/$^{24}$Mg=0.3.  Using the \citet{busso01} 
Zr yields for a 5$M_\odot$ Z=0.002 AGB star, we
estimate that a mix of 200 parts ambient material to 1 part AGB ejecta increases
the value of [Zr/Fe] by only 0.06 dex.  Therefore, at high metallicity, ejecta from AGB stars 
that have experienced hot bottom burning may
increase the ratio of $^{26}$Mg/$^{24}$Mg to the highest levels observed without producing
a detectable enhancement of $s$-process elements.  

Intermediate mass AGB stars are expected to eject
N-rich material along with the neutron-rich Mg isotopes and $s$-process elements.
HD 25329 is overabundant in N \citep{carbon87}, $s$-process elements \citep{bs94},
and $^{25,26}$Mg \citep{gl2000}.  
\citet{ventura02} calculated yields from low-metallicity AGB stars including the
effects of hot bottom burning.  They find enhancements of N by a factor of 30
despite ``conservative assumptions on the third dredge up''.  However, repeating the 
above exercise of
mixing 200 parts ambient material to 1 part AGB ejecta would not produce a significant
enhancement of N.  Since stars are known to be N-rich and overabundant in 
$^{25,26}$Mg, the models may require more realistic assumptions about the third dredge up
or a smaller dilution factor.

We have unveiled the crucial role played by AGB stars in the
chemical evolution of globular cluster NGC 6752 \citep{6752}.  This cluster
contains stars which display a significant star-to-star abundance variation in O, Na, Mg,
and Al.  At one extreme of the abundance variation, ``normal'' stars have 
elemental compositions similar to field stars at the same metallicity ([Fe/H]=$-$1.6) that are 
well explained by ejecta from metal-poor massive stars dying as 
Type II supernovae.  However, the Mg isotopic ratios found in
these ``normal'' cluster giants ($^{24}$Mg:$^{25}$Mg:$^{26}$Mg $\simeq$ 80:10:10)
exceed predictions ($^{24}$Mg:$^{25}$Mg:$^{26}$Mg $\simeq$ 98:1:1) from
massive stars.  We note that at [Fe/H]=$-1.6$, ratios of 80:10:10 match the upper
envelope in Figures \ref{fig:25mg} and \ref{fig:26mg}.
Zero metallicity AGB stars can raise the low abundances of
$^{25}$Mg and $^{26}$Mg provided by the supernovae to the high
levels observed in the normal stars.
At the other extreme of the abundance variation, ``polluted'' stars 
are underabundant in O and Mg and overabundant in Na and Al
with respect to the normal stars and have 
$^{24}$Mg:$^{25}$Mg:$^{26}$Mg $\simeq$ 60:10:30.  For [Fe/H]=$-1.6$, 
ratios of 60:10:30 greatly exceed the highest values 
in Figures \ref{fig:25mg} and \ref{fig:26mg}.
We refer to these stars as polluted because their compositions are well
explained by ejecta from AGB stars of the same metallicity as the cluster.

\subsection{Which stellar population?}

Earlier we mentioned that our sample was kinematically 
selected such that a considerable fraction have Galactic space 
velocities and Fe abundances indicative of halo or thick disk stars.  
At low metallicities, the models predict the Mg isotopes for halo stars.
At high metallicities, the predicted Mg isotopic ratios are for the thin disk whereas
our kinematically selected sample of higher metallicity stars likely contain
a mix of thick and thin
disk stars.  In effect, we may be comparing two different
stellar populations: predictions for thin disk stars and measurements
from thick disk stars.  
In Figures \ref{fig:25mg} and \ref{fig:26mg}, the solar Mg isotopic ratio 
appears rather low relative to the isotopic ratios in stars with similar metallicities.
Is this due to the sample's bias to thick disk stars?
The seven stars
observed in February 2003 were selected to be dwarfs around solar metallicity with 
thin-disk kinematics.  All seven stars had Mg isotope ratios similar to the solar 
values
(see Table \ref{tab:param}). 
This suggests that the sun is a representative member of the thin disk.

\citet{gilmore83} first measured the scale height of the Galactic thick disk and since then
many efforts have been devoted to characterizing the thick disk population.
Recent work (e.g., \citealt{fuhrmann98}, \citealt{prochaska00}, \citealt{feltzing03}, 
\citealt{mashonkina03}, and \citealt{bdp03}) indicates that below [Fe/H]=$-$0.3, thick disk
stars are overabundant in Mg ([Mg/Fe]=0.4) relative to thin disk stars ([Mg/Fe]=0.1).  
This difference (also evident for Ti, Ca, and other elements) has been attributed to the
importance of Type II supernovae relative to Type Ia supernovae in the chemical
history of the thick disk population.  The high ratios of [$\alpha$/Fe] at 
high metallicities are a defining characteristic of the thick disk population.  
Not only are these thick disk stars chemically different from the thin disk,
consideration of their space velocities also reveals systematic differences.

\begin{figure}
\epsscale{1.0}
\plotone{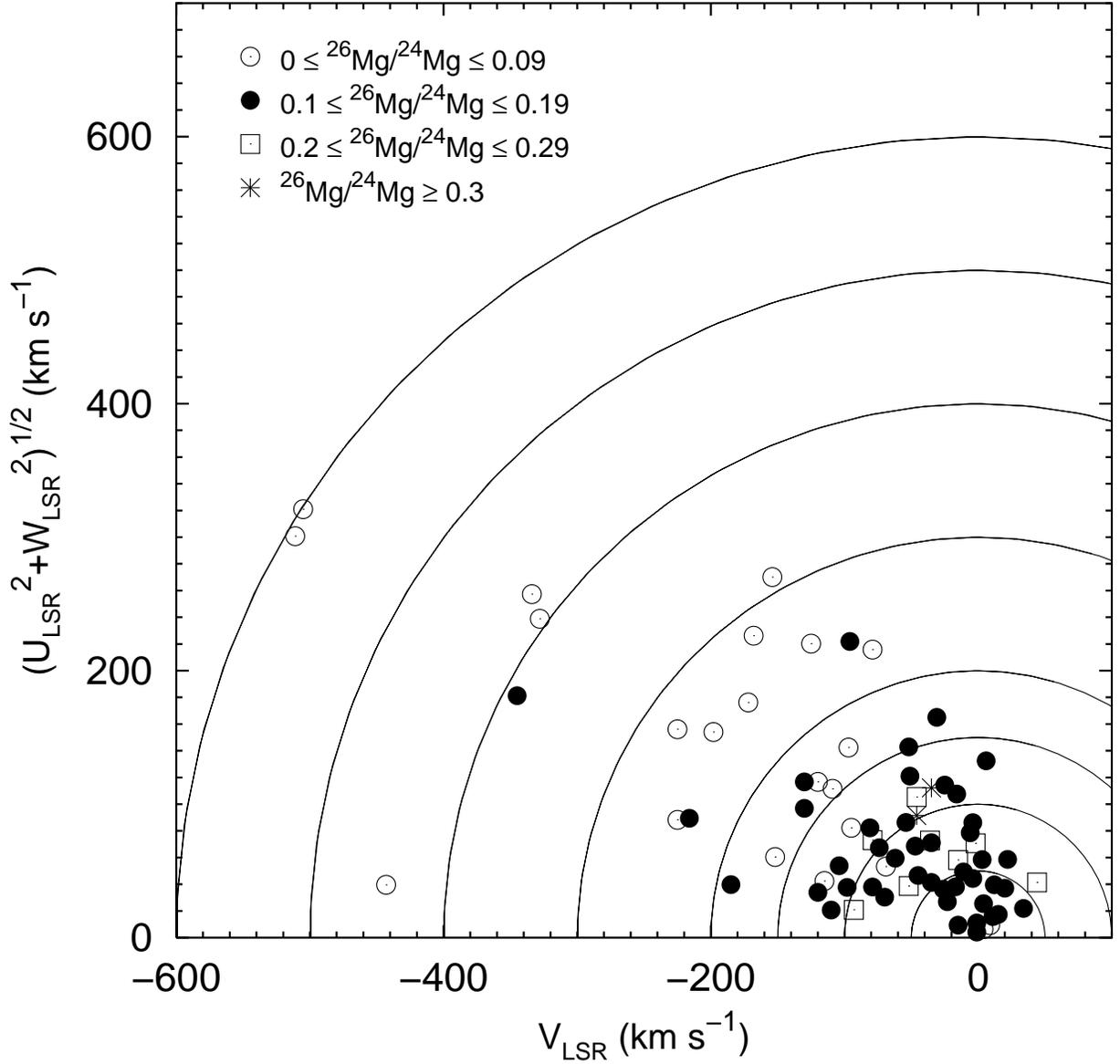}
\caption{Toomre diagram for our stars and the Gay \& Lambert sample.  The concentric circles
are lines of constant kinetic energy.  The open circles represent 
stars with $0\le^{26}$Mg$/^{24}$Mg$\le0.09$, 
the closed circles represent stars with $0.1\le^{26}$Mg$/^{24}$Mg$\le 0.19$,
the open squares represent stars with $0.2 \le^{26}$Mg$/^{24}$Mg$\le 0.29$,
and the asterisks represent stars with $^{26}$Mg$/^{24}$Mg$\ge 0.3$.
Note the absence of stars with $^{26}$Mg$/^{24}$Mg$ \ge 0.2$ and space velocities
within U$_{\rm LSR}^2$ + V$_{\rm LSR}^2$ + W$_{\rm LSR}^2 \le (50~{\rm km~s}^{-1})^{2}$.
The majority of stars lie beyond 
U$_{\rm LSR}^2$ + V$_{\rm LSR}^2$ + W$_{\rm LSR}^2 \ge (50~{\rm km~s}^{-1})^{2}$
demonstrating that have deliberately selected against stars with thin disk kinematics.
\label{fig:iso.uvw}}
\end{figure}

Figure \ref{fig:iso.uvw} shows the effect of our selection 
criteria which is biased against thin disk stars.
The Toomre diagram of our stars shows that only 15 of the 75 stars 
are likely thin disk stars with 
U$^2$ + V$^2$ + W$^2 \le (50~{\rm km~s}^{-1})^{2}$.
All stars with $^{26}$Mg/$^{24}$Mg $\ge 0.2$ lie in the range
$-150 \le$ V(km s$^{-1})$ $\le -50$ and
$0 \le$ (U$^2$+W$^2)^{1/2}$ (km s$^{-1})$ $\le 150$
which is the region of the Toomre diagram from which candidate thick disk stars 
are selected (e.g., \citealt{fuhrmann00} and \citealt{feltzing03}).

\begin{figure}
\epsscale{1.0}
\plotone{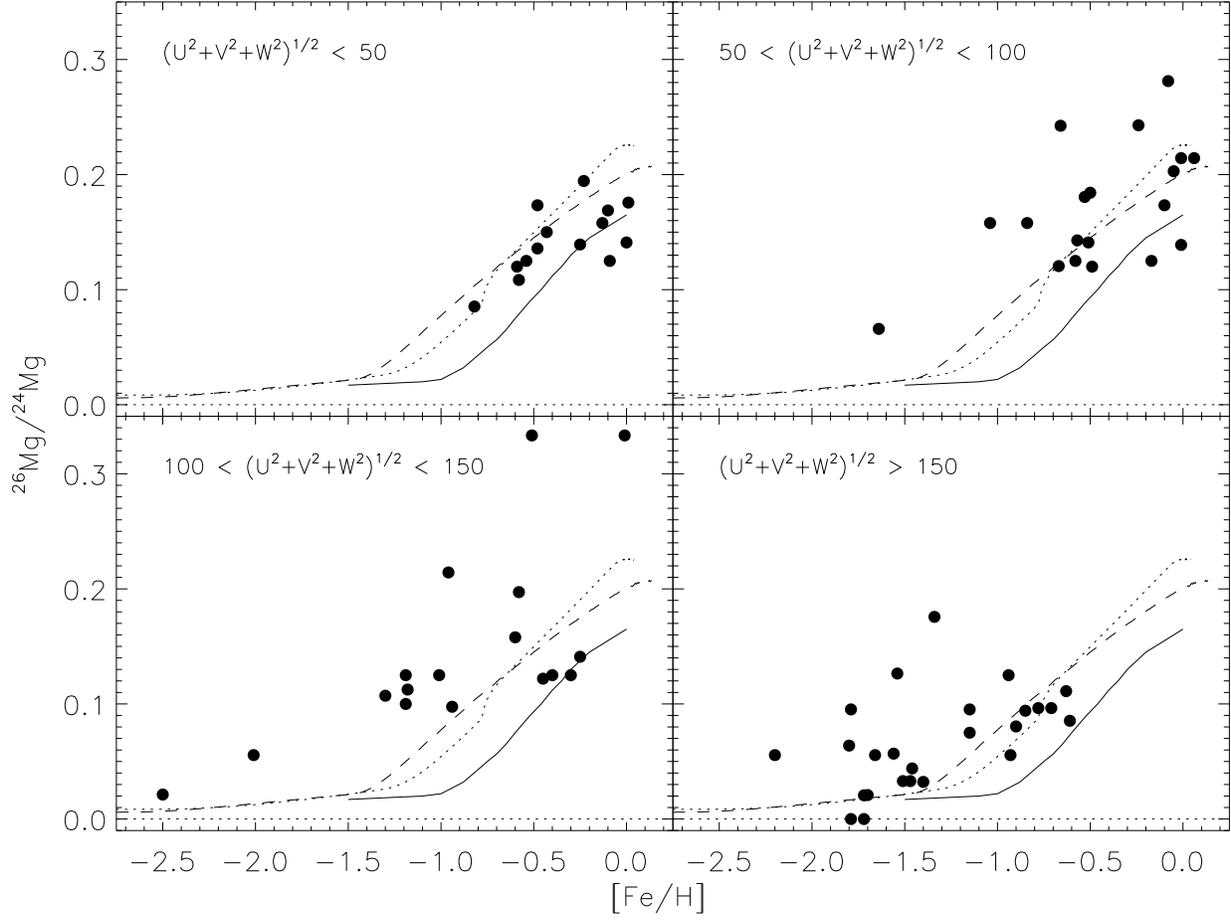}
\caption{Isotopic ratio $^{26}$Mg/$^{24}$Mg versus [Fe/H] grouped by
space velocities.  We plot (U$^2$+V$^2$+W$^2)^{1/2} \le 50$ (upper left),
$50 < ($U$^2$+V$^2$+W$^2)^{1/2} \le 100$ (upper right), 
$100 < ($U$^2$+V$^2+$W$^2)^{1/2} \le 150$ (lower left), and
(U$^2$+V$^2$+W$^2)^{1/2} > 150$ (lower right) where all velocities are in
km s$^{-1}$.  In each panel we plot the TWW1995 (solid line), GP2000 (dashed line), and ALC2001 
(dotted line) predictions.
\label{fig:iso}}
\end{figure}

In Figure \ref{fig:iso}, we separate our sample according to their kinematics and
plot the ratio $^{26}$Mg/$^{24}$Mg versus [Fe/H].  A striking trend is evident
when we consider only those stars with (U$^2$ + V$^2$ + W$^2)^{1/2} \le 50~{\rm km~s}^{-1}$.   
For these stars, the Mg isotope ratio $^{26}$Mg/$^{24}$Mg is in excellent agreement with the 
predictions.  The small scatter about the predicted curve may be entirely attributable to the
measurement uncertainties.  It should be noted that the scatter in abundances (i.e., [X/Fe]) 
at a given [Fe/H] is also very small for thin disk stars for X=C to X=Eu \citep{bdp03}.
These stars are almost certainly members of the thin disk 
based on their kinematics and metallicities.  Next we consider stars with
50 km s$^{-1} < $ (U$^2$ + V$^2$ + W$^2)^{1/2} \le 100~{\rm km~s}^{-1}$.  While some
of the sample are in good agreement with the predictions, others exceed the predictions.
The stars in this sample extend to lower metallicities than the
(U$^2$ + V$^2$ + W$^2)^{1/2} \le 50~{\rm km~s}^{-1}$ sample.  Based on the kinematics and
metallicities, these stars are a mix of thin and thick disk stars.  Now we consider stars with
100 km s$^{-1} < $ (U$^2$ + V$^2$ + W$^2)^{1/2} \le 150~{\rm km~s}^{-1}$.  A small
fraction of the sample show good agreement with the predictions with the remaining stars
exceeding the predictions.  The stars in this kinematic range exhibit the largest
spread in the Mg isotope ratios and the largest spread in [Fe/H].  Presently, the scatter in
elemental abundances (i.e., [X/Fe]) at a given [Fe/H] is not well known for the thick disk.
It is small for halo stars.  These stars are a
mix of halo and thick disk stars with the possibility that some stars may belong to the
high velocity tail of the thin disk distribution.  Finally we consider the stars with
(U$^2$ + V$^2$ + W$^2)^{1/2} > 150~{\rm km~s}^{-1}$.  About half the sample are in
good agreement with the predictions while the other half exceed the predictions.  
These stars span a smaller range in [Fe/H] than the 
100 km s$^{-1} < $ (U$^2$ + V$^2$ + W$^2)^{1/2} \le 150~{\rm km~s}^{-1}$
sample and the total spread in Mg isotope ratios is also smaller 
(the highest value belongs to the CH star BD +5 3640).  This sample 
contains halo stars and possibly some thick disk stars.

Stars we unambiguously identify as thin disk members have Mg isotope ratios in excellent
agreement with predictions with little scatter about the predicted curve.
When we consider a sample of stars that contain a mix of thin disk, thick disk,
and halo stars, the Mg isotope ratios show a large scatter including stars that
greatly exceed the predictions.  The high Mg isotopic ratios may come from thick disk stars.
Let us assume that Type II supernovae are almost entirely responsible for the iron
abundance with little, if any, contribution from Type Ia supernovae.  Under this assumption, 
Type II supernovae are responsible for the Fe and Mg.  From the
\citet{woosley95} yields, metal-rich Type II supernovae produce significant 
amounts of $^{25}$Mg and $^{26}$Mg relative to $^{24}$Mg.  So it
may be reasonable to expect high ratios of $^{25,26}$Mg/$^{24}$Mg in
thick disk stars.  It would be useful to obtain predictions of the Mg isotope
ratios from a Galactic chemical evolution model that neglects 
the contribution from Type Ia supernovae.  In such
a model, do stars with [Fe/H] $\simeq 0$ have Mg isotopic ratios 
similar to the values we measure?

\subsection{Scatter}

Attributing the scatter in Mg isotopic ratios at 
a fixed [Fe/H] to different fractions of Type II supernovae to 
AGB ejecta in the star-forming interstellar clouds 
is an appealing idea. Yet, there is an alternative which 
may be as plausible and appealing.  

The yield of $^{25}$Mg and $^{26}$Mg (relative to $^{24}$Mg) 
in Type II SN ejecta is firmly expected to be  
dependent on the stars' initial composition, as 
explained earlier. (We assume here that the composition 
and initial mass of a star are 
the key parameters. If other factors -- angular momentum, for 
example -- are relevant, the argument that follows is 
weakened.)  The yield of $^{24}$Mg is only slightly dependent on 
the initial composition of the massive stars.  
Relative to the $^{24}$Mg yield, yields for some 
elements will vary little with initial metallicity, e.g., 
Si and Ca, i.e., $\alpha$-elements.  Yields for other elements 
will be metallicity dependent, e.g., Na and Al.  

Consider now the case in which stars form in interstellar
clouds contaminated to differing degrees with Type II SN
ejecta.  If the ejecta provided to all clouds come from
stars of the same initial composition, the Mg/H ratio of the stars will be high
for stars from severely contaminated clouds and lower for those
stars from lightly contaminated clouds.  Abundance ratios such as
Mg/Si and $^{26}$Mg/$^{24}$Mg 
may be almost independent of the degree of contamination.

More likely than this 
simplest of scenarios is the case of clouds contaminated to differing 
degrees by supernovae ejecta from stars of different initial metallicity.  
The stars born in these clouds 
will show different Mg/H ratios but abundance ratios such 
as Mg/Si may vary little across the stellar sample because their respective 
yields are only weakly dependent on initial metallicity of the 
supernovaes' progenitors.  In contrast, a ratio like $^{26}$Mg/$^{24}$Mg 
comprised of species whose relative yields are metallicity dependent 
will vary across a sample of stars of the same Mg/H.  
This scenario is presumably most relevant to clouds of low initial 
metallicity and least relevant to clouds of high metallicity. 

An assessment of the applicability of this simple idea as a
partial explanation for the scatter in Figures \ref{fig:25mg} and \ref{fig:26mg}
must await the full abundance analyses of our stars. 
A few comments are offered.

For thin disk stars, the scatter in the Mg isotopic ratios is
not detectably greater than the errors of measurement.  This is
not a surprising result because the stars are among the most metal-rich of
our sample and, hence, formed from clouds contaminated by several generations
of supernovae.  More directly, elemental abundance ratios at a given [Fe/H]
show no intrinsic scatter across a large sample of
thin disk main sequence stars \citep{bdp03}.  Elements 
investigated include those whose yields from Type II supernovae
are dependent on the massive stars' initial composition.

Stars with [Fe/H] $\le -1.5$ show a spread in the Mg isotopic
ratios from the very low values predicted from metal-poor
massive stars to values (say, $^{26}$Mg/$^{24}$Mg $\sim$ 0.05)
representative of supernovae ejecta from stars more
metal-rich than the observed metallicity of our stars.  Our simple
scenario implies that the supernovae feeding these stars'
natal clouds had a composition Z $\simeq$ 0.01.
It may be possible to test this conjecture using abundance
ratios such as Na/Mg, as indicated above.  By using several
such ratios, it should be possible also to differentiate between
Mg isotopic ratios resulting from our scenario and those 
attributable to contamination of natal clouds by ejecta from
AGB stars. 

The scenario cannot, however, account for the stars with
the highest Mg isotopic ratios, say $^{26}$Mg/$^{24}$Mg $\ge$ 
0.2.  These demand either unexpectedly high isotopic
ratios from Type II SN or contamination of the star (or its natal cloud)
by ejecta from an intermediate mass AGB star.  Again, close scrutiny of
a suite of elemental abundances should be helpful.

\subsection{Non-zero ratios at low metallicity}

We claimed that 6 stars below [Fe/H]=$-1.5$ have non-zero Mg isotope ratios, 
$^{26}$Mg/$^{24}$Mg $>0.05$.  Our uncertainties in the isotope
ratios include errors in the model parameters and errors in the fits and
even allowing for these uncertainties, our Mg isotope
ratios still exceed the predictions (Figures \ref{fig:25mg} and
\ref{fig:26mg}).  At low metallicities, the
predictions from all three models are essentially identical since
they all make use of the \citet{woosley95} massive star yields.
As a consequence of primary production of the neutron-rich
Mg isotopes, the models predict a low, but non-zero, plateau in the ratios
$^{25,26}$Mg/$^{24}$Mg.
While more observations are required to confirm that the
measurements exceed the predicted plateau, two possibilities exist.
Either the supernovae yields are in error or intermediate mass stars
contribute ejecta to low metallicity gas.  

If intermediate mass stars contribute their ejecta to low metallicity gas,
then the ratio $^{26}$Mg/$^{24}$Mg will also exceed the predictions
discussed earlier.  These AGB stars may also leave another signature,
enhancements in $s$-process elements, that may be used to identify the
origin of the enhanced $^{25,26}$Mg.
Ultimately, in order to fully explore either scenario, more measurements
of Mg isotope ratios at low metallicities are essential.

\section{Concluding remarks}
\label{sec:summary}

Our 61 Mg isotopic abundance ratios confirm the decrease in
$^{25}$Mg/$^{24}$Mg and $^{26}$Mg/$^{24}$Mg with decreasing
[Fe/H].  We compared the evolution of the Mg isotopes with
predictions from models of Galactic chemical evolution
where the Mg isotopes are the product of massive stars.
The comparison highlights that, in general, our observed ratios exceed the
predictions.  Inclusion of the ejecta from AGB stars,
known to synthesize and eject the neutron-rich minor Mg isotopes,
may reconcile the observations and predictions.  There is a scatter
in the observed ratios that exceeds the measurement uncertainties
and the scatter increases with increasing metallicity.  
Our kinematically selected sample may contain a considerable fraction
of thick disk stars.
The comparison between our observed ratios and the predicted ratios for the
disk is then effectively a comparison between two different stellar
populations.  
Regardless of whether we have observed thin and/or thick disk stars, around
solar metallicity the Mg isotope ratios range from solar to more than
twice the solar value.  Mg isotope ratios in thin disk stars 
(identified kinematically) show an excellent agreement with the predictions
with little scatter about the predicted curve.

At high metallicities, are AGB stars responsible for the large
abundances of $^{25}$Mg and $^{26}$Mg?  What fraction of the stars with large
abundances of $^{25}$Mg and $^{26}$Mg are thick disk stars
devoid of products from Type Ia supernovae?
Do these stars show peculiarities in any other elemental abundances?
We have commenced an analysis of the abundances of various alpha,
iron peak, and neutron-capture elements, beginning with the stars that
have $^{26}$Mg/$^{24}$Mg $>$ 0.2.  
Only BD +5 3640 shows enhancements in $s$-process
elements and \citet{tomkin99} have already demonstrated that this is
a CH star.  Presumably thick disk stars would be readily identified by
an excess in alpha elements whereas stars contaminated by AGB
ejecta may be marked by overabundances in $s$-process elements.  At this 
preliminary stage, we do not see either signature in the stars.  Though
isotopic ratios are unaffected, uncertainties in our stellar parameters 
may mask subtle differences in elemental abundances.

Below some critical metallicity, the AGB stars will not have had time
to evolve and eject their material into the ISM.  Metal-poor
dwarfs formed from early material will then have compositions
reflecting nucleosynthesis in prior generations of massive metal-poor stars.  
Observations of isotopic ratios in the range
[Fe/H] $<-2.0$ offer the opportunity to test predicted yields
from massive metal-poor stars.  Is there a plateau in the
Mg isotopic ratios at low metallicities as predicted from
the models?  If the plateau exists, does it agree with the
predicted value?  There is evidence that our observed ratios
are higher than predictions below [Fe/H]=$-$1.5.
Further observations of Mg isotopes in the range
[Fe/H]$<-2.0$
will also show when intermediate mass AGB stars begin contributing to
the Galactic chemical evolution.  
These observations offer the chance 
to refine our understanding of stellar nucleosynthesis.
However, these cool metal-poor
stars must first be identified.  We are continuing to search for 
low luminosity cool metal-poor dwarfs.  

Not only should the search for metal-poor dwarfs be continued,
but a parallel investigation should be made of the
appropriateness of the standard method of analysis -- classical
model atmospheres, LTE etc.  One simple test would be 
to measure by the standard methods the Mg isotope ratio in
main sequence stars from an open cluster, say the Hyades.  In
such a sample, it is plausible to assume that the stars are
chemically homogeneous.  A variation of the Mg isotope ratios
with effective temperature would then signal the presence of
systematic errors afflicting the standard methods of analysis.
Then, the challenge would be to identify those errors.

\acknowledgments

We thank Amanda Karakas and Maurizio Busso for providing 
comprehensive yields from intermediate mass
AGB stars.  We thank Javier Labay and Nicolas Prantzos for providing their
predictions in electronic form.  DY thanks Amanda Karakas, Bacham Reddy, 
Carlos Allende Prieto, Gajendra Pandey, Jennifer Simmerer, John Lattanzio, and
Yeshe Fenner for many helpful discussions.  We thank the anonymous referee
for helpful comments.
DLL and DY acknowledge support from the Robert A. Welch
Foundation of Houston, Texas.
III funding support is from NASA through Hubble Fellowship 
grant HST-HF-01151.01-A from the Space Telescope Science 
Inst., operated by AURA, under NASA contract NAS5-26555.
This research has made use of the SIMBAD database,
operated at CDS, Strasbourg, France and
NASA's Astrophysics Data System.

\clearpage

\clearpage

\clearpage

\clearpage

\clearpage

\clearpage

\clearpage

\clearpage

\clearpage

\clearpage

\clearpage

\clearpage

\clearpage

\clearpage

\clearpage

\begin{deluxetable}{lcrcrcccrrrrrrr} 
\tabletypesize{\tiny}
\tablecolumns{15} 
\tablewidth{0pc} 
\tablecaption{The program stars  \label{tab:param}}
\tablehead{ 
\colhead{Name}          & \colhead{R.A.\tablenotemark{a}}     &
\colhead{Decl.\tablenotemark{b}}           &
\colhead{\teff} &
\colhead{[Fe/H]}        & \colhead{log g} &
\colhead{$\xi_t$}       & \colhead{Macro} &
\colhead{U$_{\rm LSR}$} &
\colhead{$\sigma_U$}    & \colhead{V$_{\rm LSR}$} &
\colhead{$\sigma_V$}    & \colhead{W$_{\rm LSR}$} &
\colhead{$\sigma_W$}    & \colhead{$^{24}$Mg:$^{25}$Mg:$^{26}$Mg} \\
\cline{7-14} \\
\colhead{}              & \colhead{(2000)}     &
\colhead{(2000)}     & 
\colhead{(K)} &
\colhead{}              & \colhead{(cm s$^{-2}$)} &
\multicolumn{8}{c}{(km/s)} & \colhead{}
}
\startdata 
PLX 5805\tablenotemark{c} & 000155 & 260015 & 4600 & -1.72 & 4.25 & 1.0 & 1.00 & 140 & 10 & -97 & 3 & -26 & 8 & 97:01:02 \\
HD 5098\tablenotemark{d,h} & 005241 & -240021 & 4350 & -0.43 & 1.80 & 1.7 & 4.50 & -27 & 5 & 12 & 1 & -29 & 1 & 80:08:12 \\
G 70-35\tablenotemark{d,f} & 010426 & -022159 & 4998 & -0.67 & 4.50 & 0.6 & 3.00 & 58 & 2 & 3 & 1 & 7 & 1 & 83:07:10 \\
HD 6833\tablenotemark{d,g} & 010952 & 544420 & 4400 & -0.93 & 1.10 & 1.7 & 5.00 & 131 & 4 & -198 & 2 & 81 & 7 & 90:05:05 \\
HD 9731\tablenotemark{d,h} & 013452 & -234159 & 4350 & -0.48 & 1.80 & 1.8 & 4.25 & -36 & 10 & -4 & 2 & -26 & 2 & 75:12:13 \\
BD -4 290 & 015218 & -032643 & 4900 & -0.01 & 4.50 & 0.7 & 2.00 & 70 & 10 & -2 & 1 & -10 & 5 & 70:15:15 \\
COOL 340 & 021726 & 270827 & 4800 & -0.40 & 5.00 & 0.8 & 1.25 & 34 & 6 & -98 & 17 & 16 & 7 & 80:10:10 \\
BD +4 415 & 023435 & 052647 & 4750 & -0.63 & 4.50 & 0.6 & 2.50 & 142 & 6 & -52 & 3 & -15 & 6 & 81:10:09 \\
GJ 1064a & 034702 & 412538 & 5000 & -1.15 & 4.50 & 0.6 & 1.00 & -86 & 3 & -109 & 7 & -71 & 4 & 80:14:06 \\
GJ 1064b & 034703 & 412542 & 4800 & -1.15 & 4.50 & 0.8 & 2.00 & -85 & 6 & -130 & 17 & -80 & 9 & 84:08:08 \\
GL 158 & 040315 & 351624 & 4800 & -1.79 & 5.00 & 1.0 & 1.00 & -30 & 20 & -185 & 4 & 26 & 1 & 84:08:08 \\
G 39-36 & 044807 & 330936 & 4200 & -2.50 & 4.00 & 1.0 & 1.50 & 6 & 8 & -95 & 25 & 82 & 20 & 94:04:02 \\
G 81-41 & 045505 & 454405 & 4550 & -0.96 & 4.50 & 0.6 & 2.50 & 12 & 4 & -79 & 12 & 72 & 11 & 70:15:15 \\
BD -10 1085 & 050422 & -100859 & 5100 & -0.50 & 4.50 & 0.8 & 1.50 & 14 & 3 & -70 & 4 & -27 & 1 & 76:10:14 \\
BD +19 869 & 051253 & 194320 & 4600 & -1.01 & 4.50 & 0.5 & 2.00 & 20 & 1 & -110 & 9 & -5 & 1 & 80:10:10 \\
PLX 1219\tablenotemark{c} & 052310 & 331130 & 4600 & -1.79 & 4.00 & 0.6 & 1.00 & -220 & 15 & -125 & 90 & 9 & 5 & 100:00:00 \\
G 249-37\tablenotemark{i} & 060625 & 635007 & 5000 & -0.13 & 4.50 & 0.8 & 2.75 & -26 & 1 & -23 & 1 & -7 & 1 & 76:12:12 \\
HIP 30567 & 062530 & 484340 & 4700 & -0.60 & 4.75 & 0.6 & 2.50 & 28 & 4 & -104 & 12 & -46 & 5 & 76:12:12 \\
G 103-50 & 064008 & 282712 & 4700 & -2.20 & 4.50 & 0.4 & 4.50 & 215 & 6 & -79 & 29 & 19 & 13 & 90:05:05 \\
G 88-1 & 065828 & 185949 & 4600 & -0.84 & 5.00 & 1.0 & 2.50 & 22 & 3 & -45 & 6 & 41 & 4 & 76:12:12 \\
BD -17 1716 & 065929 & -171517 & 4600 & -0.08 & 4.50 & 0.4 & 1.50 & -33 & 3 & 44 & 3 & -25 & 3 & 64:18:18 \\
HIP 33848\tablenotemark{i} & 070136 & 065537 & 5200 & 0.00 & 4.50 & 0.8 & 3.25 & 17 & 1 & 15 & 1 & 4 & 1 & 78:11:11 \\
G 87-27 & 071008 & 371634 & 5100 & -0.61 & 3.50 & 0.9 & 2.50 & 30 & 8 & -225 & 76 & -83 & 25 & 82:11:07 \\
HIP 36827\tablenotemark{i} & 072436 & -065348 & 5000 & -0.25 & 4.50 & 1.1 & 4.50 & 15 & 1 & 11 & 1 & -5 & 1 & 79:10:11 \\
G 40-5\tablenotemark{i} & 080435 & 152151 & 5200 & 0.06 & 4.50 & 0.8 & 3.75 & -38 & 1 & -52 & 1 & -7 & 2 & 70:15:15 \\
PLX 2019\tablenotemark{c} & 082941 & -014448 & 4600 & -1.72 & 4.50 & 1.0 & 2.50 & 257 & 101 & -334 & 104 & -11 & 14 & 100:00:00 \\
HIP 42145\tablenotemark{i} & 083528 & 414425 & 4800 & -0.25 & 4.50 & 0.6 & 2.50 & -67 & 1 & -74 & 3 & 8 & 1 & 78:11:11 \\
G 9-13\tablenotemark{i} & 083951 & 113122 & 5000 & -0.58 & 4.50 & 0.5 & 1.75 & 28 & 1 & -26 & 1 & -23 & 1 & 83:08:09 \\
HIP 44526\tablenotemark{i} & 090421 & -155451 & 4800 & -0.09 & 4.50 & 1.0 & 4.50 & 1 & 1 & -1 & 1 & -4 & 1 & 80:10:10 \\
LP 788-55 & 095433 & -192100 & 4700 & -0.53 & 4.50 & 0.7 & 1.50 & 86 & 18 & -4 & 6 & -5 & 4 & 72:15:13 \\
LP 429-17 & 100108 & 141842 & 4750 & -0.10 & 4.50 & 0.4 & 2.00 & 67 & 10 & -47 & 13 & -15 & 3 & 75:12:13 \\
BD +53 1395 & 101357 & 523024 & 4500 & -1.04 & 4.50 & 0.5 & 1.50 & 36 & 1 & -79 & 3 & 12 & 1 & 76:12:12 \\
BD +12 2201 & 102220 & 120845 & 4500 & -0.23 & 4.50 & 0.9 & 3.50 & -37 & 3 & -17 & 1 & -9 & 2 & 72:14:14 \\
LP 790-19 & 102607 & -175843 & 4300 & -0.51 & 5.00 & 0.6 & 1.50 & -91 & 7 & -46 & 1 & 9 & 2 & 60:20:20 \\
BD +31 2175 & 104016 & 304855 & 4800 & -0.17 & 4.75 & 1.1 & 3.50 & -70 & 4 & -35 & 2 & -13 & 3 & 80:10:10 \\
PLX 2529.1 & 105203 & -000938 & 4800 & -0.58 & 4.50 & 0.5 & 1.50 & 59 & 3 & -6 & 3 & -52 & 2 & 80:10:10 \\
BD -10 3216 & 111111 & -105703 & 4500 & -1.19 & 4.50 & 0.4 & 2.00 & -102 & 3 & -16 & 1 & 34 & 1 & 80:10:10 \\
BD +15 2325 & 112219 & 142644 & 4800 & -0.58 & 4.50 & 0.6 & 2.00 & -105 & 12 & -46 & 5 & -9 & 5 & 71:15:14 \\
LP 733-14 & 113829 & -135006 & 4700 & -0.57 & 4.75 & 0.7 & 1.50 & 5 & 3 & -35 & 11 & -41 & 10 & 77:12:11 \\
HD 103036\tablenotemark{e} & 115150 & -054544 & 4200 & -1.80 & 0.10 & 7.3 & 7.25 & -27 & 80 & -443 & 476 & -29 & 335 & 94:00:06 \\
LP 734-54 & 115754 & -094848 & 4800 & -0.66 & 4.50 & 0.4 & 2.00 & -58 & 13 & -15 & 4 & 5 & 2 & 66:18:16 \\
WOLF 1424 & 120019 & 203543 & 4600 & -1.19 & 4.50 & 0.4 & 2.50 & -8 & 4 & -120 & 25 & -33 & 6 & 80:12:08 \\
HIP 62627 & 124956 & 711139 & 4800 & -0.24 & 5.00 & 1.2 & 2.50 & 12 & 2 & -93 & 3 & -17 & 2 & 70:13:17 \\
HD 114095 & 130826 & -071830 & 4730 & -0.71 & 2.40 & 1.3 & 3.50 & -179 & 53 & -96 & 17 & 131 & 15 & 83:08:08 \\
LHS 2715 & 131857 & -030418 & 4400 & -1.56 & 4.00 & 0.6 & 1.50 & -20 & 7 & -120 & 7 & 115 & 1 & 88:07:05 \\
G 255-44 & 135536 & 740012 & 4900 & -1.30 & 4.50 & 0.7 & 1.00 & -87 & 17 & -25 & 15 & 74 & 7 & 84:07:09 \\
G 65-22 & 140144 & 085517 & 5000 & -1.66 & 4.50 & 0.4 & 2.00 & 208 & 101 & -168 & 90 & -89 & 54 & 90:05:05 \\
HIP 70152 & 142114 & 085816 & 4800 & -0.05 & 5.00 & 1.0 & 2.50 & 70 & 7 & -36 & 5 & -20 & 4 & 69:17:14 \\
BD +23 2751 & 145342 & 232043 & 4700 & -0.51 & 4.50 & 0.6 & 1.50 & -56 & 2 & -62 & 2 & 20 & 1 & 78:11:11 \\
HIP 74235 & 151013 & -162246 & 4900 & -1.51 & 4.50 & 0.8 & 1.00 & 315 & 2 & -505 & 18 & -62 & 10 & 91:06:03 \\
HD 141531\tablenotemark{e} & 154917 & 093642 & 4273 & -1.64 & 0.80 & 5.5 & 5.50 & 53 & 23 & -69 & 42 & -6 & 9 & 91:02:06 \\
G 17-25 & 163442 & -041345 & 4950 & -1.46 & 4.50 & 0.3 & 2.25 & -103 & 4 & -172 & 10 & -143 & 4 & 91:05:04 \\
G 19-25 & 172559 & -024436 & 4900 & -2.01 & 4.50 & 0.4 & 2.00 & 31 & 11 & -115 & 24 & -29 & 6 & 90:05:05 \\
BD +5 3640 & 181222 & 052404 & 4950 & -1.34 & 4.50 & 0.4 & 2.00 & 81 & 13 & -216 & 21 & 38 & 5 & 74:13:13 \\
G 23-1\tablenotemark{d,f} & 192814 & 140006 & 5200 & -0.48 & 4.50 & 0.8 & 3.75 & 2 & 1 & 20 & 1 & -37 & 2 & 81:08:11 \\
HD 232078\tablenotemark{d,g} & 193812 & 164826 & 4000 & -1.54 & 0.30 & 2.6 & 7.00 & -172 & 34 & -345 & 28 & -57 & 55 & 87:03:11 \\
BD +30 4633 & 221206 & 313341 & 4600 & -0.01 & 4.50 & 0.8 & 2.00 & 112 & 6 & -35 & 1 & -6 & 2 & 60:20:20 \\
HD 211075\tablenotemark{d,h} & 221420 & 180113 & 4350 & -0.54 & 1.50 & 1.8 & 4.50 & 11 & 2 & -1 & 2 & 1 & 3 & 80:10:10 \\
HIP 109801\tablenotemark{c}  & 221424 & -084442 & 4600 & -1.70 & 4.50 & 1.0 & 1.50 & 150 & 50 & -225 & 75 & -43 & 25 & 96:02:02 \\
ROSS 242 & 230849 & 270054 & 4700 & -0.94 & 4.50 & 0.4 & 1.50 & 70 & 14 & -130 & 16 & -67 & 30 & 80:10:10 \\
BD +28 4634 & 234510 & 293343 & 5000 & -0.30 & 4.50 & 0.7 & 2.50 & -118 & 3 & 6 & 2 & -60 & 1 & 80:10:10 \\
\enddata 

\tablenotetext{a}{hhmmss}
\tablenotetext{b}{ddmmss}
\tablenotetext{c}{These stars were taken with R = 35,000}
\tablenotetext{d}{These stars were taken with R = 120,000}
\tablenotetext{e}{These stars were taken from \citet{6752}}
\tablenotetext{f}{Stellar parameters taken from \citet{carney94}}
\tablenotetext{g}{Stellar parameters taken from \citet{pilachowski96}}
\tablenotetext{h}{Stellar parameters taken from \citet{shetrone96b}}
\tablenotetext{i}{These stars were selected to be around solar metallicity with
thin disk kinematics}

\end{deluxetable} 

\begin{deluxetable}{llcc} 
\tabletypesize{\scriptsize}
\tablecolumns{7} 
\tablewidth{0pc} 
\tablecaption{Comparison of the Mg isotopic ratios 
with the \citet{gl2000} values.  
There is a good agreement 
particularly for the ratio $^{26}$Mg/$^{24}$Mg. \label{tab:comp}}
\tablehead{ 
\colhead{Object}	& \colhead{Other name} &
\multicolumn{2}{c}{$^{24}$Mg:$^{25}$Mg:$^{26}$Mg} \\
\colhead{} 	 	& \colhead{} &
\colhead{This study} 	& \colhead{GL2000}
}
\startdata 
GJ 1064A &	HD 23439A &	80:14:06 &	78:13:09 \\
GJ 1064B &	HD 23439B &	84:08:08 &	84:08:08 \\
GL 158 &	HD 25329 &	84:08:08 &	85:08:08 \\
G 87-27 &	BD +37 1665 &	82:11:07 &	85:09:06 \\
HD 114095 &		 &	83:08:08 &	79:13:08 \\
HIP 74235 &	HD 134439 &	91:06:03 &	91:06:03 \\
\enddata 

\end{deluxetable} 

\begin{deluxetable}{lcc} 
\tabletypesize{\scriptsize}
\tablecolumns{7} 
\tablewidth{0pc} 
\tablecaption{Comparison of the Mg isotopic ratios 
with the \citet{shetrone96b} values.  
\label{tab:comp2}}
\tablehead{ 
\colhead{Object}	& 
\multicolumn{2}{c}{$^{24}$Mg:$^{25}$Mg:$^{26}$Mg} \\
\colhead{} 	 	& 
\colhead{This study} 	& \colhead{Shetrone96}
}
\startdata 
HD 5098 &   80:08:12 & 90:05:05 \\
HD 9731 &   75:12:13 & 80:10:10 \\
HD 211075 & 80:10:10 & 92:04:04 \\
HD 103036 & 94:00:06\tablenotemark{a} & 94:03:03 \\
HD 141531 & 91:02:06\tablenotemark{a} & 90:05:05 \\
\enddata 

\tablenotetext{a}{Mg isotopic ratio from \citet{6752} based on 
high resolution (R=110,000) high signal-to-noise (250 per pixel)
spectra obtained with UVES on the VLT}

\end{deluxetable}

\end{document}